\documentclass[apj,numberedappendix]{emulateapj}
\usepackage{comment} 
\usepackage{enumerate}

\shorttitle{STAR FORMATION AND ACCRETION HISTORIES AT HIGH REDSHIFT}
\shortauthors{MURPHY ET AL.}
\slugcomment{Draft version 2.2; Accepted to ApJ February 18, 2011}
\begin{document}

%\title{Star Formation and AGN Histories of 24~$\micron$ Selected Sources to {\it \MakeLowercase{z}}~$\la$2.8} 
\title{An Accounting of the Dust-Obscured Star Formation and Accretion Histories Over the Last $\sim$11~Billion Years} 

\author{E.J.~Murphy\altaffilmark{1}, R.-R. Chary\altaffilmark{1}, M.~Dickinson\altaffilmark{2}, A.~Pope\altaffilmark{2,3,6}, D.T.~Frayer\altaffilmark{4}, and L.~Lin\altaffilmark{5}}

\altaffiltext{1}{\scriptsize {\it Spitzer} Science Center, MC 314-6, California Institute of Technology, Pasadena, CA 91125; emurphy@ipac.caltech.edu}
\altaffiltext{2}{\scriptsize National Optical Astronomy Observatory, Tucson, AZ 85719}
\altaffiltext{3}{\scriptsize Department of Astronomy, University of Massachusetts, Amherst, MA 01003, USA}
\altaffiltext{4}{\scriptsize National Radio Astronomy Observatory, PO Box 2, Green Bank, WV 24944, USA}
\altaffiltext{5}{\scriptsize Institute of Astronomy \& Astrophysics, Academia Sinica, Taipei 106, Taiwan}
\altaffiltext{6}{\scriptsize {\it Spitzer} Fellow}

\begin{abstract}
We report on an accounting of the star formation and accretion driven energetics of 24~$\micron$ detected sources in the Great Observatories Origins Deep Survey (GOODS) North field.  
For sources having infrared (IR; $8-1000~\micron$) luminosities $\gtrsim3\times 10^{12}~L_{\sun}$ when derived by fitting local spectral energy distributions (SEDs) to 24~$\mu$m photometry alone, we find these IR luminosity estimates to be a factor of $\sim$4 times larger than those estimated when the SED fitting includes additional 16 and 70~$\mu$m data (and in some cases mid-infrared spectroscopy and 850~$\mu$m data).  
This discrepancy arises from the fact that high luminosity sources at $z\gg0$ appear to have far- to mid-infrared ratios, as well as aromatic feature equivalent widths, typical of lower luminosity galaxies in the local Universe.  
Using our improved estimates for IR luminosity and AGN contributions, we investigate the evolution of the IR luminosity density versus redshift arising from star formation and AGN processes alone.  
We find that, within the uncertainties, the total star formation driven IR luminosity density is constant between $1.15 \la z \la 2.35$, although our results suggest a slightly larger value at $z\ga2$.  
AGN appear to account for $\la$18\% of the total IR luminosity density integrated between $0\la z \la 2.35$, contributing $\la$25\% at each epoch.  
Luminous infrared galaxies (LIRGs; $10^{11} \leq L_{\rm IR} < 10^{12}~L_{\sun}$) appear to dominate the star formation rate (SFR) density along with normal star-forming galaxies ($L_{\rm IR} < 10^{11}~L_{\sun}$) between $0.6 \la z \la 1.15$.  
Once beyond $z\ga2$,  the contribution from  ultraluminous infrared galaxies (ULIRGs; $L_{\rm IR} \geq 10^{12}~L_{\sun}$) becomes comparable with that of LIRGs.   
Using our improved IR luminosity estimates, we find existing calibrations for UV extinction corrections based on measurements of the UV spectral slope typically overcorrect UV luminosities by a factor of $\sim$2, on average, for our sample of 24~$\mu$m-selected sources; 
accordingly we have derived a new UV extinction correction more appropriate for our sample.  
\end{abstract}

\section{Introduction}
Precisely quantifying the stellar mass assembly over cosmic time is critical for understanding galaxy formation and evolution.  
While the first studies which set out to characterize the cosmic star formation history had to rely on observations of the emergent (rest-frame) UV emission, requiring significant corrections for extinction by dust, 
%\citep[e.g.,][]{pm99}, 
the advent of space-based infrared observatories have allowed us to measure the dust-obscured star formation activity directly.  
%The resolution of the cosmic infrared background using ISO \citep[e.g.,]{de99,so00,hd01}, {\it Spitzer} \citep[e.g.,][]{el04,el05,kc06,cp07,bm09}, and now {\it Herschel}, has uncovered populations of dusty, starbursting galaxies with infrared (IR; $8-1000$) luminosities ranging between $10^{10}~L_{\odot} \la L_{\rm IR} \la 10^{13}~L_{\sun}$.  
The resolution of the cosmic infrared background using {\it ISO} \citep[e.g.,][]{de02}, {\it Spitzer} \citep[e.g.,][]{hd06}, and now {\it Herschel} \citep[e.g.,][]{sb10}, has uncovered populations of dusty, starbursting galaxies with infrared (IR; $8-1000$) luminosities ranging between $10^{10}~L_{\odot} \la L_{\rm IR} \la 10^{13}~L_{\sun}$.  
These cosmologically important luminous ($10^{11}~L_{\odot} \leq L_{\rm IR} < 10^{12}~L_{\sun}$) and ultraluminous ($L_{\rm IR} \geq 10^{12}~L_{\sun}$) infrared galaxies, LIRGs and ULIRGs respectively, appear to dominate the star formation rate (SFR) density in the Universe between redshifts of $1 \la z \la 3$  \citep[e.g.,][]{ce01, el05,kc07}.   
%Thus, a more direct way to measure the evolution of star formation activity is by directly measuring the dust obscured component.  

Many of these studies rely on deep 24~$\mu$m imaging and the use of local spectral energy distribution (SED) libraries to extrapolate the evolution of the total infrared luminosity with redshift.  %\citep[e.g.,][]{el05,kc07}.  
In doing so, large (factors of $\sim$10) bolometric corrections are invoked to scale up the rest-frame mid-infrared flux densities and estimate the total SFR per unit co-moving volume.  
Using the rest-frame mid-infrared wavelengths, specifically near $\sim$8~$\mu$m, can be problematic due to the presence of broad spectral features attributed to polycyclic aromatic hydrocarbon (PAH) molecules.  
While the rest-frame 8~$\mu$m emission correlates with the total IR luminosity of galaxies in the local Universe, albeit with a large amount of scatter \citep[e.g.,][]{dd05,jd07,la07} and systematic departures for low-metallicity systems \citep[e.g.,][]{ce05,sm06}, there are hints that this correlation may breakdown at increasing redshift.    
In a recent study, \citet{ejm09a} has shown that galaxies in the redshift range between $1.4 \la z \la 2.6$ exhibit PAH equivalent widths that are large compared to local galaxies of similar luminosity.    
This translates into an overestimate of IR luminosities from rest-frame mid-infrared photometry which has been noted for subsets of $z\sim2$ galaxies \citep[e.g.,][]{cp07,jr08,de10,rn10}.  
%{\bf Other studies have also found that rest-frame mid-infrared derived IR luminosities overestimate the actual IR luminosity of sources at $z\sim2$ through stacking \citep[e.g.,]{cp07}, observations of lensed systems \citep[e.g.,][]{jr08}, and now using new {\it Herschel} data \cite[e.g.,][]{de10,rn10}. }

An additional uncertainty arises from the fact that a fraction of the light emitted by these IR-bright galaxies may also be associated with embedded active galactic nuclei (AGN).  
The separation of the AGN and star-forming component of a galaxy's IR output requires knowledge of its full SED, which is difficult to measure for a large number of high redshift systems.  
One technique to get at this separation of AGN and star formation activity is through the use of mid-infrared spectroscopy where one 
% of galaxy samples selected by their mid-infrared flux densities provides a powerful technique to isolate these various selection effects among infrared luminous sources.
decomposes the spectra into aromatic feature and continuum components \citep[e.g.,][]{as07,ap08,ejm09a}.   
%one can get a handle on the relative fraction of power being emitted by AGN activity within such sources relative to the output related to star formation \citep[e.g.,][]{as07,ap08}.  
For a heterogenous sample of IR-bright galaxies this technique has revealed that the contribution of AGN to the total IR luminosity output can span a large range, and is, on average, non-negligible being roughly $\sim$30\% \citep{ejm09a}.  

In this paper we 
%build on the work of \citet{ejm09a} by using 
use deep 70~$\mu$m data from the Far-Infrared Deep Legacy (FIDEL; P.I.: M. Dickinson) survey to provide improved estimates for the total IR luminosities of 24~$\micron$-selected sources in the Great Origins Observatories Deep Survey-North \citep[GOODS-N;][]{md03} field out to $z\approx2.8$.  
Using these improved values, along with an empirical relation for the fractional AGN contribution to the total IR luminosities of the 24~$\mu$m-selected sources, we show how the IR luminosity density evolves separately for AGN and star formation versus redshift.  
In doing so, we also look at how the contribution from populations of normal galaxies ($L_{\rm IR} < 10^{11}~L_{\sun}$), LIRGs, and ULIRGs to the SFR density varies as a function of lookback time.  
% for all galaxies, LIRGs, and ULIRGs and look for variation in the IR-radio correlation out to $z\sim3$.  
%Using these improved values 
%While doing so, 
%In this paper we also compare UV, IR, and radio based SFRs among the sample.    
%We investigate how the frequency of ``mid-infrared" excess sources,  defined as galaxies having mid-infrared based SFRs in excess to those derived from extinction corrected UV estimates \citep{ed07a,ed07b} changes compared to when IR luminosities are based on 24~$\micron$ photometry alone.  
 
The paper is organized as follows:  
In $\S$2 we describe the sample properties and observations.  
IR luminosity estimates through SED fitting, along with a description of how we estimate contributions from AGN, are discussed in $\S$3.    
%The results of our comparison between the IR and radio properties of the sample are presented in $\S$4.  %
%In $\S$5 we investigate how the frequency of ``mid-infrared" excess sources,  defined as galaxies having mid-infrared based SFRs in excess to those derived from extinction corrected UV estimates in the $BzK$ redshift range \citep{ed07a,ed07b} changes compared to when IR luminosities are based on 24~$\micron$ photometry alone. 
%We also search for variations in the IR-radio correlation as a function of redshift.  
A look into the applicability of standard UV extinction estimates from the UV spectral slope for our sample of 24~$\mu$m-selected sources is given in $\S$4.  
A discussion of the results on the evolution of the IR luminosity density versus redshift is given in $\S$5, along with an explicit description of the uniqueness of this study.    
Finally, in $\S$6, we summarize our conclusions.  

\section{Sample Selection and Multiwavelength Photometry}
\begin{figure*}
\plotone{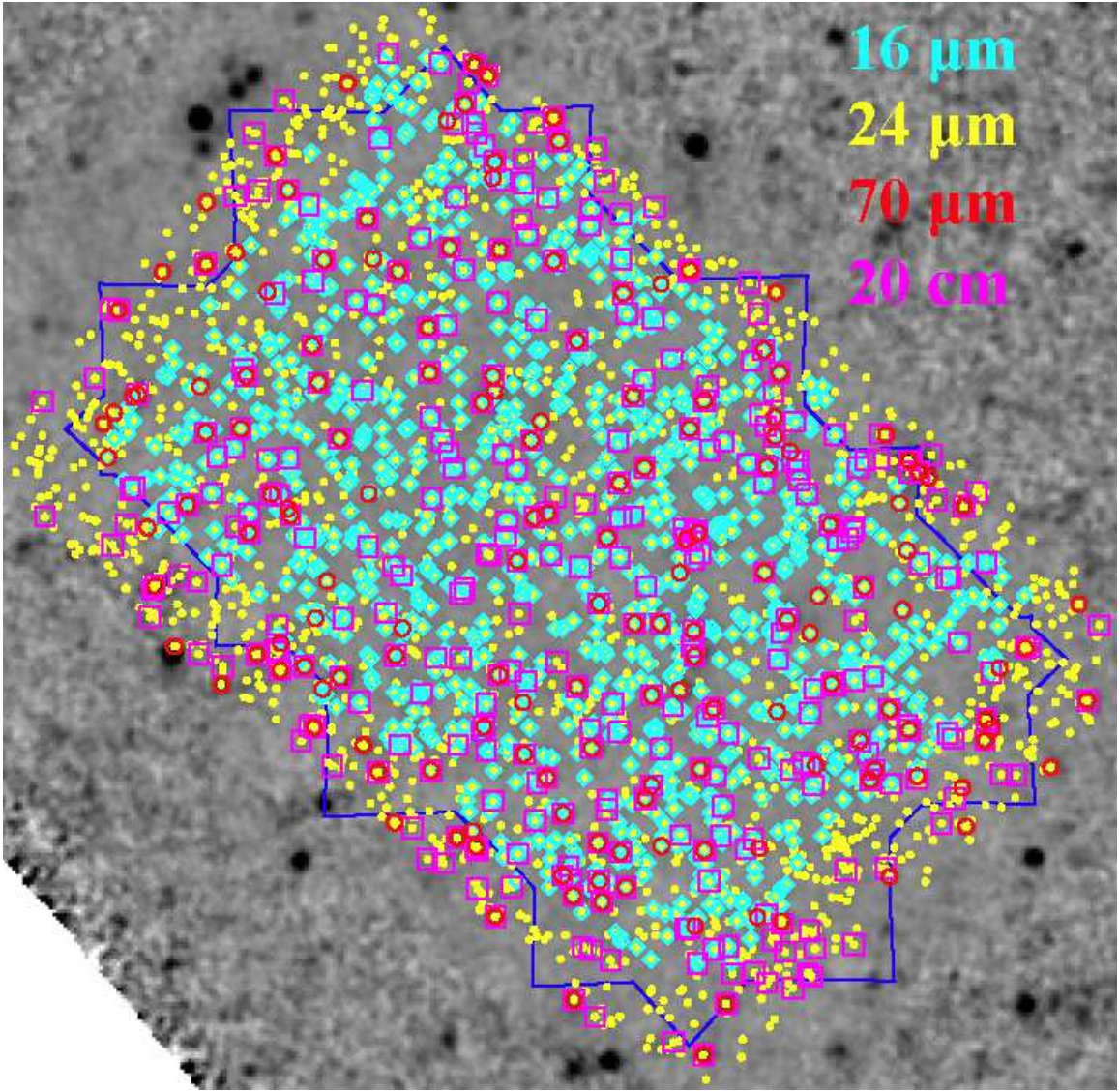}
\caption{The location for each of the 2664 24~$\mu$m detected sources overlaid on the MIPS~70~$\micron$ map of the GOODS-N field.  
Corresponding detections at 16~$\mu$m, 70~$\micron$ and 20~cm are shown; 
note that the areal coverage of the 16~$\mu$m map is smaller than the extent of the 24~$\mu$m observations.  
%Source detections at 16, 24, 70~$\micron$ and 20~cm overlaid on the MIPS~70~$\micron$ map of the GOODS-N field.  
The blue outline shows the $\approx$160 arcminute$^2$ ACS areal coverage over which our analysis of the IR luminosity evolution versus redshift is conducted.  
%{\bf SHOW ACS COVERAGE OUTLINE}
\label{fig-1}}
\end{figure*}

%We present a study on the infrared characteristics of galaxies selected at 24~$\micron$ in the GOODS-N field.  
The GOODS-N field is located around the {\it Hubble} Deep Field-North at $12^{\rm h}36^{\rm m}55^{\rm s}$, $+62\degr 14\arcmin 15\arcsec$ (J2000).  
In Figure \ref{fig-1} we overlay source detections on the 70~$\micron$ image of GOODS-N \citep{df06} and provide a summary of the source detections at 16, 24, and 70~$\mu$m, as well as in the optical and hard-band ($2.0-8.0$~keV) X-rays in Table \ref{tbl-1} (see Appendix A for details of existing radio data).  %, as well as at 20~cm, in Table \ref{tbl-1}.  
%In the last column of Table \ref{tbl-1} we also give 
The total number of sources detected in all 3 infrared bands is 112.  %, and in the radio and infrared (82).  
The multiwavelength data used here are the same that were used in the analysis of the mid-infrared spectroscopic sample presented in \citet{ejm09a}.  

\begin{deluxetable}{cccccc}
\tablecaption{Summary of Detections within ACS Coverage \label{tbl-1}}
\tablewidth{0pt}
\tablehead{
\colhead{} & \colhead{16~$\micron$} & \colhead{24~$\micron$} &
\colhead{70~$\micron$} & \colhead{Optical ($z_{850}$)} & \colhead{Hard X-Ray} %& \colhead{All IR}
}
\startdata
  $N_{\rm ACS}$& 1122  & 2196  &  133  &  1884 & 156 %& 112  
\enddata
\tablecomments{Out of a total of 2664 24~$\mu$m detected sources.}
\end{deluxetable}

\begin{figure}
\plotone{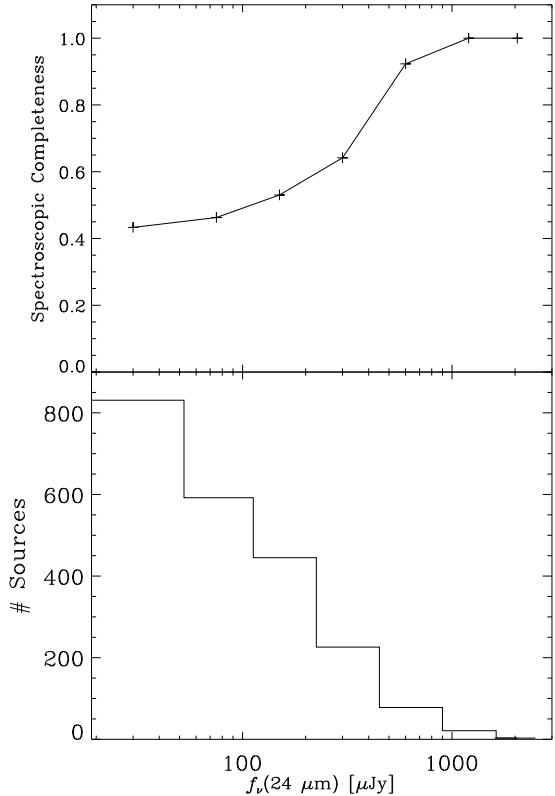}
\caption{In the top panel the spectroscopic completeness for the 24~$\mu$m sources included in the areal coverage of the ACS imaging is plotted against the 24~$\mu$m flux density.  
The completeness is differential, illustrating the fraction of sources with spectroscopic redshifts per each 24~$\mu$m flux density bin.  
The bottom panel shows total number counts of 24~$\mu$m sources per flux density bin.  
%The spectroscopic completeness for the 1111 24~$\mu$m sources included in the areal coverage of the ACS imaging plotted against the 24~$\mu$m flux density.  
%The completeness is cumulative; the line represents the fraction of galaxies brighter than $f_{\nu}(24~\mu$m) for which a spectroscopic redshift was measured.  
\label{fig-2}}
\end{figure}

\begin{figure}
\plotone{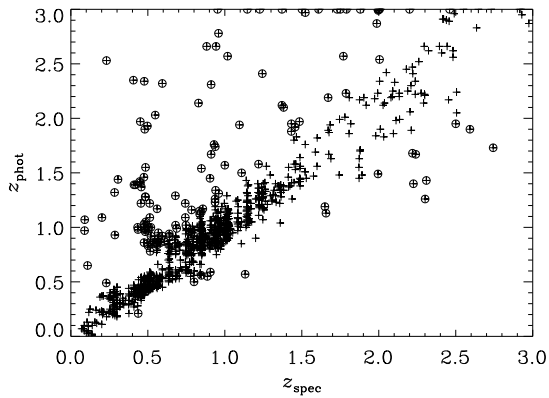}
\caption{
A comparison between the photometrically recovered redshifts to the spectroscopic values.  
The distribution of relative errors $dz = (z_{\rm phot} - z_{\rm spec})/(1 + z_{\rm spec})$ has a median of $\approx-0.005\pm0.186$.  
Outliers (i.e., those sources having $|dz| > 0.15$) are identified with open circles.  
Excluding these sources reduces the median and dispersion of the distribution of relative errors to $\approx0.003\pm0.058$.  
%difference between these two samples is $\Delta \tilde{z} \approx -0.006 \pm 0.29$.  
\label{fig-3}}
\end{figure}

\begin{figure}
\plotone{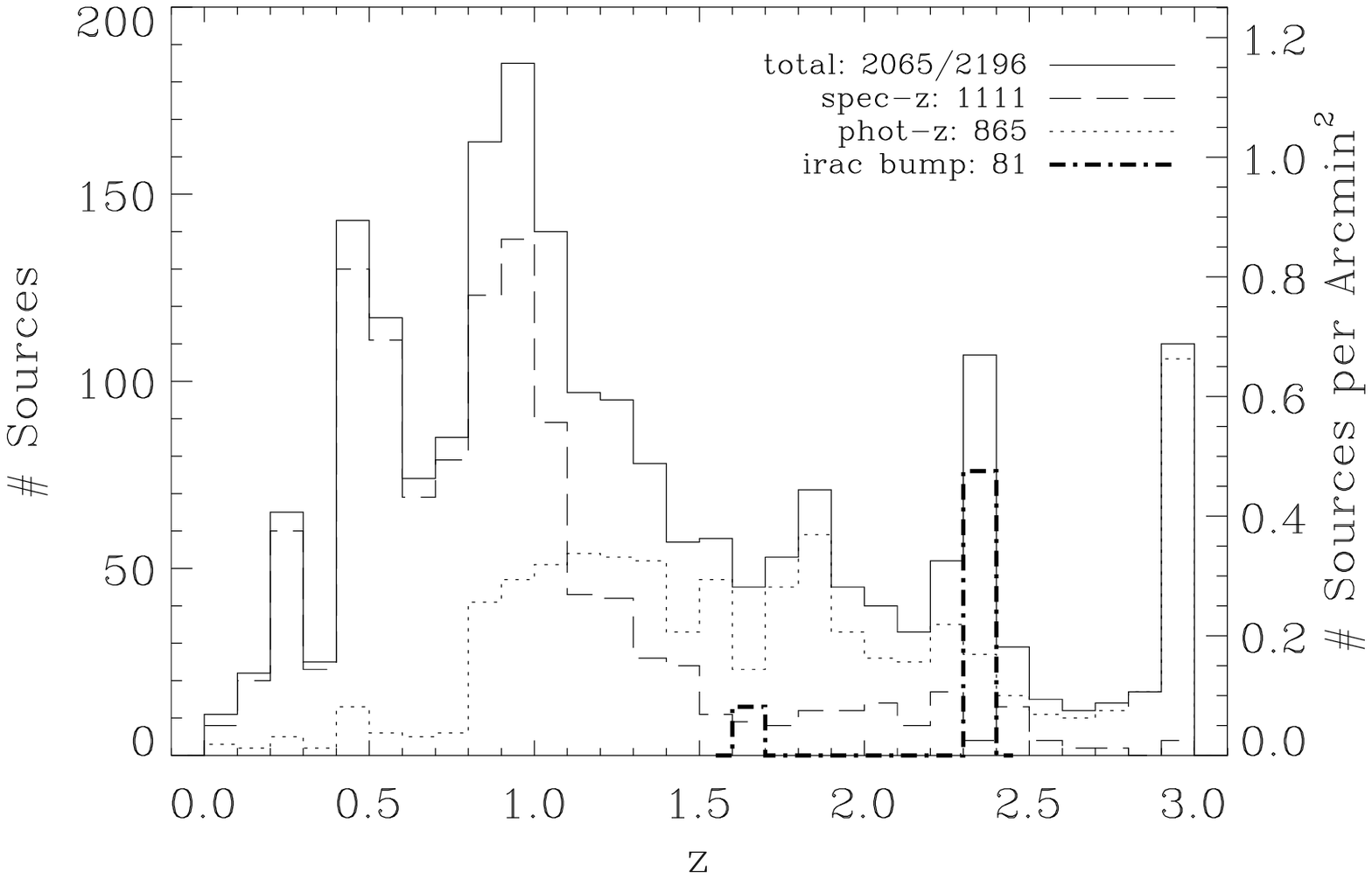}
\caption{
Histogram of the redshift distribution for 24~$\mu$m sources included in the ACS areal coverage.  
The maximum allowed photometric redshift is set to $z_{\rm phot} = 3$; such sources were not used in our analysis of the evolution of the infrared luminosity density versus redshift.  
There are 6 sources with $z_{\rm spec} \geq 3$ not shown.  
\label{fig-4}}
\end{figure}

\subsection{Mid- and Far-infrared  {\it Spitzer} Imaging}
$\it Spitzer$ observations at 16, 24, and 70~$\micron$ of the GOODS-N field were taken as part of various programs.  
%Infrared imaging of the GOODS-N field was obtained at 16, 24, and 70~$\micron$ using the $Spitzer$ Space Telescope.  
Observations at 24~$\micron$ using the Multiband Imaging Photometer for {\it Spitzer} \citep[MIPS;][]{gr04} were taken as part of the GOODS Legacy program (PI: Dickinson) and reach an rms of $\sim$5~$\mu$Jy \citep[see][]{rc07}.
The catalog was created using IRAC prior positions and sources having a 24~$\micron$ flux density greater than 20~$\mu$Jy and a signal-to-noise ratio (S/N) greater than 3 were considered to be detections leading to a total number of 2664 detections;   
%The resulting upper limit at 24~$\mu$m is 
a 24~$\mu$m flux density of 30~$\mu$Jy corresponds to a differential completeness of $\approx$80\% \citep{bm09}.  
The 24~$\micron$ flux density among these sources spans a range between $\sim$$20-4320~\mu$Jy, more than a factor of $\sim$200; the median flux density is $\sim$70~$\mu$Jy.  
Calibration uncertainty at 24~$\micron$ is $\sim$10\%.  

%Since we are interested comparing infrared luminosties estimated by SED template fitting for comparison with other star formation diagnostics as well as to look for possible evolution in the FIR-radio correlation with redshift, exact redshifts are critical for a proper analysis.  
%We therefore limit our analysis to those galaxies for which we have firm spectroscopic redshifts.  
Of the 2664 24~$\micron$ detections a total of 2196 sources are included within the optical ({\it Hubble} ACS; see $\S$2.3) areal coverage of the GOODS-N field with 1111 spectroscopic redshifts available (\citealp{jc00,gw04,bcw08}; D. Stern et al. 2010, in preparation).  
The differential spectroscopic completeness as a function of the 24~$\mu$m flux density is plotted in Figure \ref{fig-2}, along with the corresponding 24~$\mu$m number counts.
Among these sources having spectroscopic redshifts, 1105 are at $z < 3$ having a median value of $\approx$0.845 (see Table \ref{tbl-2} for a breakdown using 6 redshift bins).  
%for 1113 sources, and 1107 are less than $z\sim3$ having a median value of $\sim$0.845.  
%\citep[][; D. Stern et al. 2008, in preparation]{jc00,gw04}.  
%The spectroscopic completeness plotted as a function of 24~$\mu$m flux density for those 24~$\mu$m sources overlapping with the optical coverage is shown in Figure \ref{fig-2}.  
The 24~$\micron$ flux densities of these sources span a range from $\sim$$20-2440~\mu$Jy (i.e., a factor of $\sim$120), with a median value of  $\sim$80~$\mu$Jy

Imaging at 16~$\micron$ was carried out using one of the two peak-up array cameras for the Infrared Spectrograph for {\it Spitzer} \citep[IRS;][]{jh04}.  
Details of the imaging program and catalog can be found in \citet{ht05,ht10}.  
The areal coverage of the 16~$\micron$ map was slightly smaller than that of the GOODS-N 24~$\micron$ map (see Figure \ref{fig-1}).  
The rms noise of these data is $\sim$6~$\mu$Jy and only sources having S/N ratios greater than 5 were considered to be firm detections leading to a total of 1297 sources detected.  
Of these, 1154 have 24~$\micron$ counterparts (1122 within the ACS coverage), while 775 of these also had spectroscopic redshifts; 769 were at $z<3$.  
Among these 769 sources, their 16~$\micron$ flux densities span a range from $\sim$$30-1420~\mu$Jy (i.e., a factor of nearly $\sim$50) with a median value of 95~$\mu$Jy.  
The calibration uncertainly at 16~$\micron$ is $\sim$10\% and all non-detections were assigned the upper limit value of 30~$\mu$Jy.  

Deep MIPS 70~$\micron$ observations centered on, and extending beyond, the GOODS-N field were performed as part of two separate imaging programs; GO-3325 \citep[PI:][]{df06} and the Far Infrared Deep Extragalactic Legacy project (FIDEL; PI: Dickinson).  
The typical point-source noise of the 70~$\micron$ map is $\sim$0.55~mJy \citep{df06} which is roughly $\sim$1.6 times larger than the confusion noise level of $\sigma_{\rm c} = 0.35 \pm 0.15$~mJy \citep{df09}.   
%Details on the data processing can be found in \citet{df06}.  
%The full field extends well beyond the 24~$\micron$ coverage and 
Details of the data processing can be found in \citet{df06}.  
The cataloged sources were considered detections if their 70~$\micron$ flux density was $>~2$~mJy and they had a S/N$~>6$;  these criteria correspond to a differential completeness of  $\sim$80\% \citep{bm09} and a total of 167 sources are detected, 164 of which had 24~$\micron$ counterparts (133 within the ACS coverage).    
Of the 164 sources matched with a 24~$\mu$m counterpart, 123 also had spectroscopic redshifts, all of which were at $z<3$.  
The 70~$\micron$ flux densities range between $\sim$$2-19~$mJy, spanning nearly an order of magnitude, and have a median value of 3.7~mJy.  
Sources not meeting the detection criteria were assigned the upper limit value of 3~mJy.   
The calibration uncertainty at 70~$\micron$ is $\sim$20\%.  
A summary of the total number source detections at 16, 24, and 70~$\micron$ within the ACS areal coverage can be found in Table \ref{tbl-1}.  
% and without 24~$\micron$ counterparts, can be found in Table \ref{tbl-1}.  

\subsection{X-ray Imaging}
GOODS-N was observed using the {\it Chandra} X-ray Observatory for a 2~Ms exposure \citep{da03}.    
The full band ($0.5-8.0$~keV) and hard band ($2.0-8.0$~keV) on-axis sensitivities are $\sim$$7.1\times10^{-17}$ and $1.4\times10^{-16} {\rm erg~cm^{-2}~s^{-1}}$, respectively.   
Of the 2196 24~$\micron$ detected sources included in the ACS areal coverage, 223 were matched with an X-ray counterpart reported by \cite{da03}.  
%, while 38 positions only corresponded to upper-limits.  
%261 were matched with an X-ray counterpart reported by \cite{da03}, 38 of which were reported as upper-limits.  
A total of 156 of these sources have firm detections in the hard band.  

The full band X-ray flux spans more than 3 orders of magnitude, ranging between $3-6950\times10^{-17}~{\rm erg~cm^{-2}~s^{-1}}$, with a median flux of $87\times10^{-17}~{\rm erg~cm^{-2}~s^{-1}}$.  
The hard-band X-ray detected sources span a smaller range in flux, spanning between $9-4230\times10^{-17}~{\rm erg~cm^{-2}~s^{-1}}$, with a median flux of $135\times10^{-17}~{\rm erg~cm^{-2}~s^{-1}}$.  

\subsection{Optical and Near-Infrared Imaging}
%An $\approx$160 arcminute$^2$ region of the 
The $\approx$160 arcminute$^2$ GOODS-N field has been imaged extensively at optical wavelengths using the ACS camera on the {\it Hubble} Space Telescope in the following four filters: $B_{435},~V_{606},~ i_{775}, ~{\rm and}~z_{850}$ \citep{mg04}.  
In our analysis we use the catalogs for version 1 of the ACS data obtained from MAST.  
Additional ground-based imaging of GOODS-N in the $U$-band was obtained using the prime-focus MOSAIC camera on the KPNO Mayall 4~m telescope.   
A subset of these data were described in \citet{mg04} and \citet{cap04};  the present analysis uses a version of the $U$-band data that is has approximately twice the exposure time.
The $U$-band photometry was measured by detecting sources in the Subaru $R$-band image of \citet{cap04} and measuring $U$-band fluxes through matched apertures.
%Additional ground-based imaging of GOODS-N in the $U$-band was obtained using the prime-focus MOSAIC camera on the KPNO Mayall 4~m telescope \citep{mg04, cap04}.  
%The 3-$\sigma$ upper limits of the $UBViz$ photometry are 0.027, 0.048, 0.048, 0.091, and 0.131~$\mu$Jy, respectively.  
%Sources were considered detections in the $UBViz$ bands if they had a S/N~$>3$ and a flux density larger than 0.027, 0.048, 0.048, 0.091,and 0.131~$\mu$Jy, respectively.  
Sources were considered detections in the $UBViz$ bands if they had a S/N~$>3$, corresponding to a typical detection limit of 0.050, 0.035, 0.024, 0.035,and 0.026~$\mu$Jy, respectively.  
%was $>~2$~mJy and they had a S/N$~>6$.  
%The 3-$\sigma$ detection limits of the $UBViz$ photometry are 0.050, 0.035, 0.024, 0.035,and 0.026~$\mu$Jy, respectively.  

%We additionally make use of CFHT WIRCAM near-infrared (NIR) $JK$ data taken for GOODS-N (L. Lin et al. 2010, in preparation).  
GOODS-N has also been observed at near-infrared (NIR) wavelengths with the Wide-Field Near Infrared Camera (WIRCAM) at the CFHT, including $K_s$-band observing programs from Hawaiian and Canadian observing programs, and $J$-band data from a Taiwanese program.   
WIRCAM has a field of view covering $21\farcm 5 \times 21\farcm 5$, fully encompassing the GOODS-N ACS, IRAC and MIPS areas.
Here we use reductions of the WIRCAM data from Lin et al.\ (2010, in preparation), consisting of 27.4 hours of integration time at $J$ and 31.9 hours at $K_s$.   
The images have seeing with FWHM~$\approx 0\farcs7$.  
Sources were considered detections at $J$ and $K_s$ if they had a S/N~$>3$ and a flux density larger than 0.312 and 0.456~$\mu$Jy, respectively.  
%0.027, 0.048, 0.048, 0.091,and 0.131~$\mu$Jy, respectively.  
%and reach 3$\sigma$ detection limits of 0.239 and 0.235~$\mu$Jy at $J$ and $K_s$, respectively.   
An independent analysis of the WIRCAM $K_s$-band data has recently been published by \citet{whw10}.  
%Wang et al.\ (2010).  

IRAC data at 3.6, 4.5, 5.8, and 8.0~$\mu$m were taken as part of the GOODS {\it Spitzer} Legacy program (M. Dickinson et al. 2010, in preparation).  
The formal 3$\sigma$ limits for an isolated point source are 0.079, 0.137, 0.867, and 0.951~$\mu$Jy for channels 1,2,3, and 4, respectively.  
However, in practice, the signal to noise will depend on the degree of crowding with other nearby sources.  

\subsection{Inclusion of Photometric Redshifts}
The 24~$\mu$m detections were cross-matched against the ACS $BViz$, CFHT $JK_s$, and the 3.6 and 4.5~$\mu$m IRAC catalogs to provide photometric input for the Z-PEG code \citep{dl02} to compute photometric redshifts.  
%Using the Z-PEG code \citep{dl02} we compute photometric redshifts for each 24~$\mu$m detection using the ACS $BViz$ data, CFHT $JK_s$ data, and the 3.6 and 4.5~$\mu$m IRAC data.  
Only sources having detections (i.e. S/N ratio $> 3$) in 4 or more of these wavebands are considered to have reliable photometric redshifts.  
A maximum redshift value of 3 was used while running Z-PEG.  % as this led to slightly better results.  %assigning photometric redshifts.  
The distribution of relative errors $dz = (z_{\rm phot} - z_{\rm spec})/(1 + z_{\rm spec})$ among the 1050 sources having spectroscopic and photometric redshifts has a median and dispersion of $\approx -0.005\pm0.186$ (Figure \ref{fig-3}).  
Applying an outlier criterion of $|dz| > 0.15$, 
%$|(z_{\rm phot} - z_{\rm spec})| > 0.15(1 + z_{\rm spec})$, 
which excludes 139 ($\approx$13\%) of the sources, the median and dispersion of the relative errors are reduced to $\approx 0.003\pm0.058$.   

Using all 1050 sources (i.e., including $|dz| > 0.15$ outliers), we assign uncertainties to the photometric redshifts by taking the standard deviation of the difference between the spectroscopic and photometric redshifts ($\sigma_{\Delta z}$) within each of the last 5 redshift bins listed in Table \ref{tbl-2} resulting in uncertainties of $z \pm0.33$, 0.27, 0.28, 0.46, and 0.46, respectively.    
These values, in addition to the uncertainties on the input flux densities in the SED fitting, were used in a Monte Carlo approach when estimating uncertainties on IR luminosities  derived with photometric redshifts (see $\S$3.1.1).  
%The median difference between the 1020 sources for which a spectroscopic and photometric redshift was available, both of which are below a value of 2.8 (i.e., are redshift range of interest) is $\approx -0.006\pm0.29$ (Figure \ref{fig-3}).  

For an additional 81 sources not having enough broad-band detections to extract a reliable photometric redshift using Z-PEG, we use their peak in the IRAC channels to place them into a redshift bin based on the expected location of the 1.6~$\mu$m stellar bump.     
Taking the updated \citet{cb03} stellar templates, which include a revised prescription for stars on the asymptotic giant branch \citep[AGB; see][]{gb07}, extincted by an $A_{V}=1$, it is found that the 1.6~$\mu$m stellar bump should shift into IRAC channels 2 and 3 for redshift bins of $1.3 \la z \la 1.9$ and $1.9 \la z \la 2.8$, respectively.  
For these sources, the center of the redshift bins are used when calculating IR luminosities (i.e., $z \approx 1.6$ and 2.35, respectively).  
Uncertainties in these redshifts were estimated by taking the standard deviation of the difference between the spectroscopic and IRAC-based photometric redshifts where possible, resulting in uncertainties of $z \approx 1.6\pm0.47$ and $z\approx 2.35\pm0.69$.   
The choice of an $A_{V}= 1$ was for simplicity; we note that by increasing the assumed extinction to an $A_{V} = 3$, the center of the redshift bins decrease to $z \approx 1.44$ and 2.19, resulting in IR luminosities which are $\approx$20\% and 35\% smaller, respectively, which is within our estimated errors.      
%$1.6 \la z \la 1.9$ and $1.9 \la z \la 2.8$, respectively.  

As with the IR luminosities derived using photometric redshifts above, uncertainties in the IRAC-based photometric redshifts were used to assign an additional uncertainty to the IR luminosity through a standard Monte Carlo technique.  
The redshift distribution for the 2065 (out of 2196) 24~$\mu$m-selected sources included in the ACS areal coverage, for which we have a spectroscopic or photometric redshift, is shown in Figure \ref{fig-4}.  
The inclusion of these photometric redshifts results in a redshift completeness of $\approx$94\%.  
We note here that photometric redshifts are only used for our analysis of the evolution of the IR luminosity density versus redshift.  
%All other analyses of the SEDs and flux ratios of individual galaxies are based entirely on spectroscopically confirmed redshifts.  

%\section{Data Analysis} 
\section{Estimating Star Formation and AGN Activity} 
In the following section we briefly describe our methods for calculating infrared (IR;  $8-1000~\micron$) luminosities ($L_{\rm IR}$) for each 24~$\micron$ detected source.  
% having a spectroscopic redshift.  
In addition, we describe a method to estimate the AGN contribution to each IR luminosity.  %, as well as calculate IR, radio, and UV (1500~\AA) based SFRs.    
IR luminosities are used to estimate corresponding SFRs following the conversions given in \citet{rck98}.   
% as described in \citet{ejm09a}.  
%The determination of IR and UV (1500~\AA) based SFRs, along with the method used for estimating UV extinction corrections, is also presented.  
The analysis presented here closely follows that of \citet{ejm09a}, where a more detailed description can be found; 
in Appendix A we present a similar comparison of IR, radio, and UV (1500~\AA) based SFRs, while in Appendix B we demonstrate that the bolometric and AGN corrections given in \citet{ejm09a} are applicable to the present study of 24~$\mu$m-selected galaxies.    
% is warranted.  
%compare the mid-infrared properties of the \citet{ejm09a} to the entire 24~$\mu$m-selected sample to demonstrate that the application of their bolometric and AGN corrections to the present study is warranted.  
All calculations are made assuming a Hubble constant of 71~km~s$^{-1}$, and a standard $\Lambda$CDM cosmology with $\Omega_{\rm M} = 0.27$ and $\Omega_{\Lambda} = 0.73$.   
Throughout the paper average properties are measured by taking a median rather than a mean.    

\subsection{Estimating IR luminosities of Galaxies}
In the following subsection we describe our methodology for deriving estimates of the IR luminosities among the sample.    
We also describe how we account for the fractional contribution of AGN to the IR luminosity of each source.  

 \begin{deluxetable}{ccc}
\tablecaption{Sources per $z$ Bin \label{tbl-2}}
\tablewidth{0pt}
\tablehead{
\colhead{} & 
\colhead{$N_{24~\micron}$} & 
\colhead{$N_{70~\micron}$} 
}
\startdata
 \cutinhead{Spectroscopic Redshifts}
 0.0$~\leq~z~<~$ 0.4&      111&       30\\
 0.4$~\leq~z~<~$ 0.7&      310&       43\\
 0.7$~\leq~z~<~$ 1.0&      338&       28\\
 1.0$~\leq~z~<~$ 1.3&      176&       12\\
 1.3$~\leq~z~<~$ 1.9&       90&        6\\
 1.9$~\leq~z~<~$ 2.8&       76&        3\\
\cutinhead{Spectroscopic and Photometric Redshifts}
 0.0$~\leq~z~<~$ 0.4&      123&       31\\
 0.4$~\leq~z~<~$ 0.7&      333&       43\\
 0.7$~\leq~z~<~$ 1.0&      430&       28\\
 1.0$~\leq~z~<~$ 1.3&      333&       15\\
 1.3$~\leq~z~<~$ 1.9&      349&        6\\
 1.9$~\leq~z~<~$ 2.8&      362&        4
\enddata
\tablecomments{Only sources within the ACS areal coverage are considered.  There are 2 sources with spectroscopic redshifts outside of the ACS areal coverage, as well as 4 additional sources  with spectroscopic redshifts between  $2.8 \leq z < 3.0$.}  
\end{deluxetable}

\begin{figure*}
\plottwo{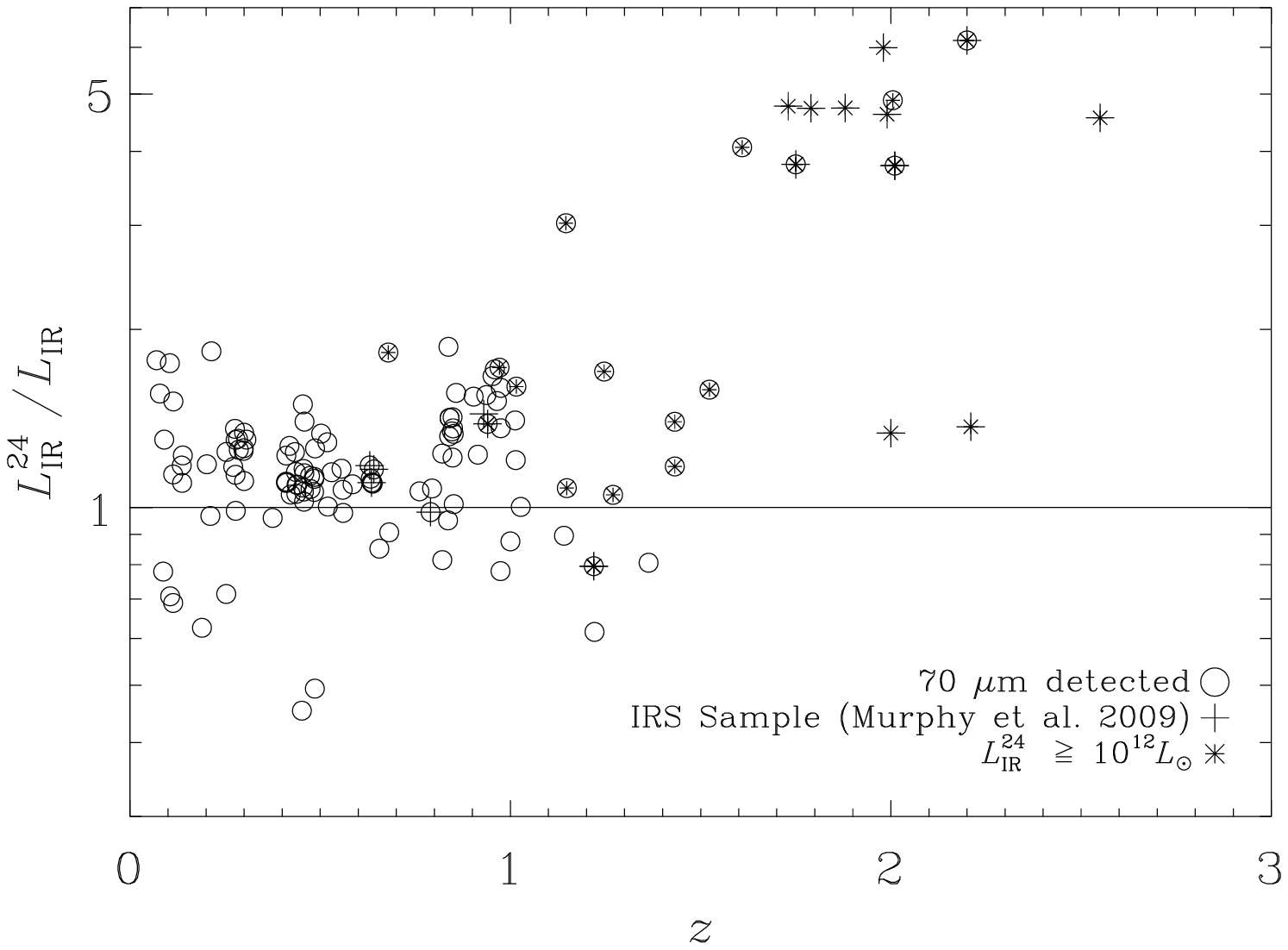}{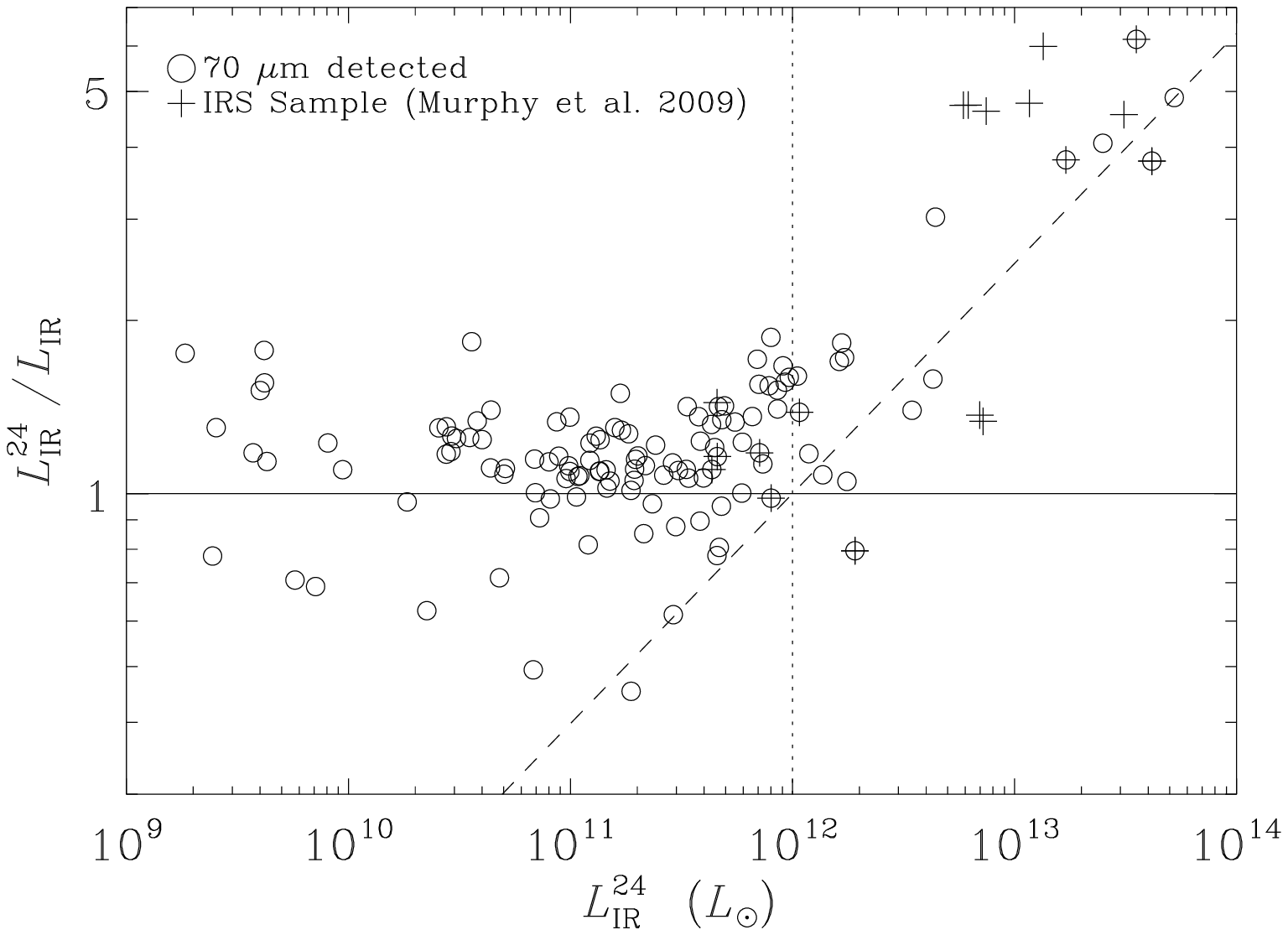}
\caption{
In the left panel we plot the ratio of the 24~$\mu$m-derived IR luminosity to those estimated by SED template fitting with the 16, 24, and 70~$\micron$ photometry for all 70~$\mu$m detections (open circles) versus redshift.  
Also included is the ratio of the 24~$\mu$m-derived to the best-fit IR luminosities for the mid-infrared spectroscopic sample of \citet{ejm09a} (plus symbols). 
In the right panel we plot the same ratios against the 24~$\mu$m-derived IR luminosity.  
ULIRGs ($L_{\rm IR} \geq 10^{12}~L_{\sun}$) are identified by asterisks.  
The vertical dotted line separates ULIRGs from non-ULIRGs and the dashed line is the ordinary least squares fit to the ULIRGs given by Equation \ref{eq-l24cor}.  
%The solid horizontal line in each panel indicates a ratio of unity.  
\label{fig-5}}
\end{figure*}

\subsubsection{Determining IR Luminosities from SED Fitting \label{sec-sed}}
%Infrared luminosities are extrapolated as described in \citet{ejm09a}; 
We fit the photometric data (i.e., the 16, 24, and 70~$\micron$ flux densities where available) with the SED templates of \citet{ce01} and then integrate between $8-1000~\micron$.  
% \citep{ejm09a}.  
The SED libraries of \citet{ce01} are used since they have recently been found to exhibit 24/70~$\micron$ flux density ratios which are better matched to the average observed for galaxies at $z\sim1$ compared to the \citet{dh02} or \citet{gl03} templates \citep{bm09}.  
The best-fit SEDs are determined by a $\chi^{2}$ minimization procedure in which the amplitude (luminosity) and shape (temperature) of SED templates are allowed to scale freely.  
Fitting errors are estimated by a standard Monte Carlo approach using the photometric uncertainties of the input flux densities, as well as redshift uncertainties for sources having only photometric redshifts.    
In the cases where only upper limit values are available for the photometric data, they are not incorporated into the formal $\chi^{2}$ minimization but are used to reject fits having associated flux densities greater than the upper limit.  
These IR luminosities are identified by $L_{\rm IR}^{16, 24, 70}$.    

For sources without far-infrared information (i.e., those sources not detected at 70~$\mu$m), which constitutes the majority of sources, $L_{\rm IR}^{16, 24, 70}$ values are estimated by averaging the IR luminosities from the best-fit \citet{ce01} and \citet{dd01} SEDs.  
%for which the \citet{dd01} templates have a luminosity normalization \citep{dm06}.   
%{\bf The luminosity normalization to the \citet{dd01} templates is described in \citep{m06}.  }
Since the \citet{dd01} SED templates are not normalized by luminosity, we do so using a local trend between IRAS colors and IR luminosity as the \citet{dd01} SEDs are described by a single parameter (i.e., $f_{\nu}(60~\micron)/f_{\nu}(100~\micron)$) family.      % \citep[e.g.,][]{dm06}  }
Due to the non-detection at wavelengths close to the peak of the far-infrared emission, the IR luminosities of such sources may be considered as upper limits. 
We note that for a heterogeneous sample (i.e., SMGs, AGNs, and optically faint sources) of 22 galaxies spanning the same redshift range being probed here, IR luminosities derived in this manner were found to be $\sim1.0 \pm 0.5$ times those estimated from SED fitting with additional mid-infrared spectra and (for 18 of the 22 galaxies) 70 and/or 850~$\micron$ photometry \citep{ejm09a}.  
  
%Similar to \citet{ejm09a}, 
%We are interested in seeing to what extent the mid-infrared excesses reported by \citet{ed07a,ed07b} can be solely attributed to incorrect bolometric corrections from SED fitting rather than the presence of an embedded AGN.  
%Therefore, 
We additionally calculate IR luminosities by fitting SEDs with the 24~$\micron$ photometry alone.  
This is also done by fitting both the \citet{ce01} and \citet{dd01} SED template libraries independently, and averaging their integrated $8-1000~\micron$ luminosities.  
The associated IR luminosity is designated as $L_{\rm IR}^{24}$, and is, on average, within $\la$5\% of those derived from fitting either set of SED templates.  
The difference between the resultant $L_{\rm IR}^{24}$ from the two sets of SED templates is characterized as a systematic error. 

\begin{deluxetable*}{ccccccc}
\tablecaption{Star Formation and Accretion Powered IR Luminosity Density Evolution \label{tbl-3}}
\tablewidth{0pt}
\tablehead{
\colhead{} & 
\colhead{} & \colhead{} & \colhead{} &
\colhead{}  & \colhead{LIRGs}  & \colhead{ULIRGs} \\
\colhead{} & 
\colhead{$\rho_{\rm IR}$} & \colhead{$\rho_{\rm SFR}^{\dagger}$} & \colhead{$\rho_{\rm AGN}^{\ddagger}$} &
\colhead{$L_{\rm IR} < 10^{11}~L_{\sun}$}  & \colhead{$10^{11}~L_{\sun} \leq L_{\rm IR} < 10^{12}~L_{\sun}$}  & \colhead{$L_{\rm IR} \geq 10^{12}~L_{\sun}$} \\
\colhead{} & 
\colhead{($10^{8}~L_{\sun}~{\rm Mpc}^{-3}$)} & \colhead{($M_{\sun}~{\rm yr}^{-1}~{\rm Mpc}^{-3}$)} & \colhead{($10^{8}~L_{\sun}~{\rm Mpc}^{-3}$)} & 
\colhead{(\%$^{*}$)} & \colhead{(\%$^{*}$)} & \colhead{(\%$^{*}$)}
}
\startdata 
 0.4$~\leq~z~<~$ 0.7&   4.78$\pm$0.31  & 0.062$\pm$ 0.005  &1.21$\pm$0.14  & 63  & 35  &  2\\
 0.7$~\leq~z~<~$ 1.0&   7.31$\pm$0.41  & 0.098$\pm$ 0.006  &1.62$\pm$0.28  & 47  & 49  &  4\\
 1.0$~\leq~z~<~$ 1.3&   8.62$\pm$0.76  & 0.122$\pm$ 0.012  &1.59$\pm$0.37  & 47  & 46  &  7\\
 1.3$~\leq~z~<~$ 1.9&   9.30$\pm$1.28  & 0.135$\pm$ 0.022  &1.48$\pm$0.50  & 39  & 41  & 20\\
 1.9$~\leq~z~<~$ 2.8&  14.39$\pm$3.45  & 0.207$\pm$ 0.059 &2.44$\pm$1.56  & 23  & 35  & 42
\enddata
\tablecomments{$^{\dagger}$\(\rho_{\rm SFR}(M_{\sun}~{\rm yr}^{-1}~{\rm Mpc}^{-3}) = 1.73\times10^{-10}[\rho_{\rm IR}(L_{\sun}~{\rm Mpc}^{-3}) - \rho_{AGN}(L_{\sun}~{\rm Mpc}^{-3})]\). 
$^{\ddagger}$These values for the AGN IR luminosity density may be considered as upper limits (see $\S$3.1.3).  
$^{*}$Percentage of the SFR density.}
\end{deluxetable*}

\subsubsection{Empirical Correction for 24~$\micron$-derived IR Luminosities \label{sec-lir24corr}}
\citet{ejm09a} have shown that IR luminosities derived by fitting local SEDs with 24~$\mu$m photometry alone typically overestimate the true IR luminosity by an average factor of $\sim$5 among sources having $z > 1.4$ and $L_{\rm IR}^{24}$ values $\ga 3\times10^{12}~L_{\sun}$.  
We use that result to derive a empirical correction for $L_{\rm IR}^{24}$.    
In the left and right panels of Figure \ref{fig-5} we plot the ratio of 24-$\micron$-derived to best-fit IR luminosities ($L_{\rm IR}^{16, 24, 70}$) for all of the 70~$\micron$ detected sources having spectroscopic redshifts, versus redshift and 24~$\micron$-derived IR luminosities, respectively.    
Also included in each panel is the ratio of 24-$\micron$-derived to best-fit IR luminosities for 18 galaxies presented by \citet{ejm09a}; 
the IR luminosities among these galaxies were fit using 70 and/or 850~$\micron$ photometry and are considered fairly well determined.  
For the nine 70~$\micron$ detected sources which overlap between the IRS and larger GOODS-N samples, we use the IR luminosity reported by \citet{ejm09a} because those estimates include the mid-infrared spectroscopic data in the fits.  

The trend reported by \citet{ejm09a} appears to hold.    
The 24~$\micron$-derived IR luminosities of galaxies having redshifts between $1.4 \la z \la 2.6$ and $L_{\rm IR}^{24} \ga 3\times 10^{12}~L_{\sun}$ are overestimated by an average factor of $\sim$$4.6\pm1.7$.  
Galaxies at lower redshifts, having $L_{\rm IR}^{24} < 10^{12}~L_{\sun}$, have 24-$\micron$-derived IR luminosities which are consistent  with our best-fit determinations using additional longer wavelength data.  
These results are also consistent with recent {\it Herschel} findings confirming the reported overestimates in 24~$\mu$m-derived IR luminosities by \cite{ejm09a} \citep[i.e.,][]{de10,rn10}.    

While trends exist between the overestimates in the bolometric correction for both redshift and 24~$\mu$m-derived IR luminosity (see right panel of Figure \ref{fig-5}), we believe the latter to be the dominant parameter as the trend between the ratio of 24~$\mu$m-derived to best-fit IR luminosity versus 24~$\mu$m-derived IR luminosity appears tighter than versus redshift; 
the residual dispersion in the fit between the ratio of 24~$\mu$m-derived to best-fit IR luminosity and redshift (for $z\ga1.3$) is $\sim$40\% larger than the fit versus luminosity, where $L_{\rm IR}^{24} \geq 10^{12}~L_{\sun}$.    
% appears more pronounced in the right panel of FIgure \ref{fig-5}.  
Furthermore, individual lensed Lyman break galaxies (LBGs) at $z\sim2.5$, having IR luminosities $\la 10^{12}~L_{\sun}$, appear to show bolometric corrections that are similar to those of local galaxies \citep{bs08, gw08,ag09}.  

Consequently, we correct the 24~$\micron$-derived IR luminosities using this information.  
For galaxies having $L_{\rm IR}^{24} < 10^{12}~L_{\sun}$ we do not apply a correction given that the median ratio of $L_{\rm IR}^{24}$ to the best-fit IR luminosity is $1.1\pm0.27$.  
For galaxies having $L_{\rm IR}^{24} > 10^{12}~L_{\sun}$, we apply an empirical correction by fitting these sources with an ordinary least square regression such that, 
%\begin{equation}  
%\label{eq-l24cor}
%\log \left(\frac{L_{\rm IR}^{\rm 24,corr}}{L_{\sun}}\right) \approx 
%\Bigg\{
%\begin{array}{ll}
%\log\left(\frac{L_{\rm IR}^{24}}{L_{\sun}}\right), & L_{\rm IR}^{24} < 10^{12}~L_{\sun} \\
%(0.59\pm0.04)\log\left(\frac{L_{\rm IR}^{24}}{L_{\sun}}\right)-(4.9\pm0.5) & L_{\rm IR}^{24} \geq 10^{12}~L_{\sun}.
%\end{array}
%\end{equation}
\begin{eqnarray}  
\nonumber
\label{eq-l24cor}
\log \left(\frac{L_{\rm IR}^{\rm 24,corr}}{L_{\sun}}\right) = &  & \\
\Bigg\{ &
\begin{array}{ll}
\log\left(\frac{L_{\rm IR}^{24}}{L_{\sun}}\right), & L_{\rm IR}^{24} < 10^{12}~L_{\sun} \\
(0.60\pm0.04)\log\left(\frac{L_{\rm IR}^{24}}{L_{\sun}}\right)+ \\
	(4.8\pm0.5), & L_{\rm IR}^{24} \geq 10^{12}~L_{\sun}.
\end{array}
\end{eqnarray}
For the cases where these IR luminosity estimates exceed those derived using the additional 16 and 70~$\mu$m data, we take the minimum value between these two values as the best estimate for the true IR luminosity of each source such that, 
\begin{equation}
L_{\rm IR} = \min(L_{\rm IR}^{\rm 24,corr}, L_{\rm IR}^{16, 24, 70}).
\end{equation}
The fraction of cases for which $L_{\rm IR}^{\rm 24,corr}$ was much larger than $L_{\rm IR}^{16, 24, 70}$ (i.e., $L_{\rm IR}^{\rm 24,corr}/L_{\rm IR}^{16, 24, 70} > 2$) occurred for $\sim$2\% of the sample.  
%This condition does not affect our results significantly as cases for which $L_{\rm IR}^{\rm 24,corr}/L_{\rm IR}^{16, 24, 70} > 2$ occur for only $\sim$2\% of the sample.  
Lastly, for the subsample of 22 sources included in \citet{ejm09a}, we set $L_{\rm IR}$ to their best-fit IR luminosities.  

\begin{figure}
\plotone{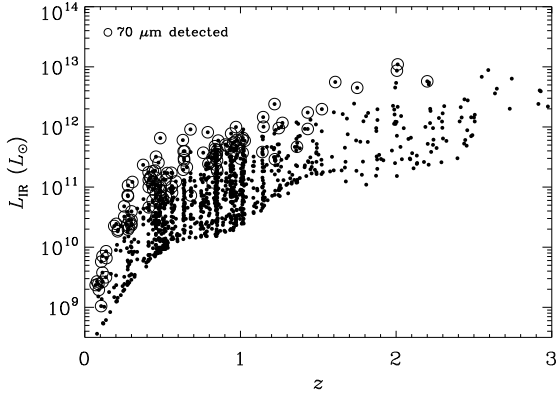}
%\plottwo{f3a.eps}{f3b.eps}
\caption{
The best-fit IR luminosities for each of the 1107 24~$\micron$ detections having spectroscopic redshifts of $z<3$.  
Sources detected at 70~$\mu$m are identified by an open circle.  
\label{fig-6}}
\end{figure}

The best-fit IR luminosities are plotted in Figure \ref{fig-6} as a function of redshift for all 1107 24~$\micron$ detected sources having spectroscopic redshifts of $z<3$.  
%; this includes two more sources than included in the ACS coverage.  
%These values for IR luminosity are considered to be our best estimate for the true IR luminosity of each source.   
The values range from $2.5\times10^{8}-1.1\times10^{13}~L_{\sun}$, spanning more than 4 orders of magnitude, with a median of $7.2\times10^{10}~L_{\sun}$.  
Limiting these to galaxies having firm 70~$\micron$ detections, the median is increased by a factor of $\sim$2.5 to $1.8\times10^{11}~L_{\sun}$ and the range decreases to  $1.0\times10^{9}-1.1\times10^{13}~L_{\sun}$.  
%shrinks by a factor of $\sim$2, spanning from $1.2\times10^{9}-1.4\times10^{13}~L_{\sun}$.  
%Instead, looking at 
On the other hand, the range of the 24~$\micron$ derived luminosities, spans more than 5 orders of magnitude from $2.7\times10^{8}-6.4\times10^{13}~L_{\sun}$, with a median value of $7.6\times10^{10}~L_{\sun}$, roughly the same as the median of the best-fit IR luminosities.

%\subsubsection{Determining AGN Fractions of IR Luminosities}
\subsubsection{Constraints on AGN Fractions of IR Luminosities}
We make a statistical correction to the IR luminosities of each source by assigning a fractional contribution from AGN based on an empirical trend between AGN luminosity and the difference between the 24~$\micron$-derived and true IR luminosities \citep{ejm09a} such that 
%We assign AGN contributions to the total IR luminosity of each source based on an empirical trend between AGN luminosity and the difference between the 24~$\micron$-derived and true IR luminosities \citep{ejm09a} such that 
%Using Equation~10 of  \citet{ejm09a}, which relates the amount of IR luminosity arising from an AGN to the difference between the 24~$\micron$-derived and true IR luminosities derived using data from the mid-infrared through the submillimeter, we assign the AGN luminosities to our sources such that, 
\begin{equation}
\label{eq-agnfrac}
%\log\left(\frac{L_{\rm IR}^{\rm AGN}}{L_{\sun}}\right) = 
%		(0.50 \pm 0.04) \log \left(\frac{L_{\rm IR}^{24} - 
%		L_{\rm IR}}{L_{\sun}}\right) + (5.61 \pm 0.52).
\log\left(\frac{L_{\rm IR}^{\rm AGN}}{L_{\sun}}\right) = 
		(0.73 \pm 0.08) \log \left(\frac{L_{\rm IR}^{24}}{L_{\sun}}\right) + (2.5 \pm 1.0).
\end{equation}
This relation relies on AGN luminosity estimates derived by first decomposing mid-infrared spectra into star formation (PAH template) and AGN (hot dust emission) components.  
Since hot dust associated with vigorous star formation may also contribute to the mid-infrared continuum emission along with AGN, our estimates for the AGN luminosity could be considered upper limits.    
However, this conclusion assumes that our choice of AGN template (i.e., Mrk~231) is appropriate for all sources, which may not be the case.  
Furthermore, it is currently unclear as to whether the far-infrared luminosity of Mrk~231 is powered primarily by AGN \citep[e.g.,][]{la07,sv09,jf10} or star formation \citep[e.g.,][]{ds98,df03}.
If the latter is true, than that would again argue for the AGN IR luminosity estimates being upper limits. 
Correspondingly, the SFR density that we have derived would be increased by this amount.
However, there is no reason to believe that these high-$z$ sources do not contain AGN which contribute significantly to powering the observed dust emission.   
Galaxies with SFR of $\sim 20~M_{\sun}~{\rm yr}^{-1}$ occur in galaxies with stellar masses of $\sim$10$^{10}$\,M$_{\sun}$ \citep[][]{ed07b} which should therefore harbor black holes of mass $\sim$10$^{7-8}$\,M$_{\sun}$ if the black hole-stellar mass relation holds at $z\sim$1 to 2 \citep{rjm06}.  
Thus, our AGN estimates may be considered upper limits, but in reality it is challenging to definitively quantify this without a precise knowledge of the AGN SED.  
%Furthermore, it is currently controversial as to whether Mrk~231 is powered primarily by AGN \citep[e.g.,][]{la07,sv09,jf10} or star formation \citep[e.g.,][]{ds98,df03} processes.  
%If the latter is true, than that would again argue for the AGN IR luminosity estimates being upper limits.  
%However, there is no reason to believe that these high-$z$ sources do not contain AGN which will power a significant amount dust emission, leading to a SED that is similar to Mrk~231 independent of what dominates the energetics in that source.  
%Thus, these estimates may be considered upper limits, but in reality it is hard to definitively quantify this.  

By subtracting these values from our best estimates for the true IR luminosity, we can quantify the amount of IR luminosity arising from star formation alone such that $L_{\rm IR}^{\rm noAGN} = L_{\rm IR} - L_{\rm IR}^{\rm AGN}$.    
For the cases in which this empirical correction leads to AGN contributions that are larger than a source's estimated IR luminosity, we assume the source is completely powered by an AGN and set  \(L_{\rm IR}^{\rm AGN} = L_{\rm IR}\).

\subsection{The Evolution of the IR Luminosity Density versus Redshift}
\begin{figure*}
\begin{center}
%\scalebox{1.15}
\scalebox{1.0}
{\plotone{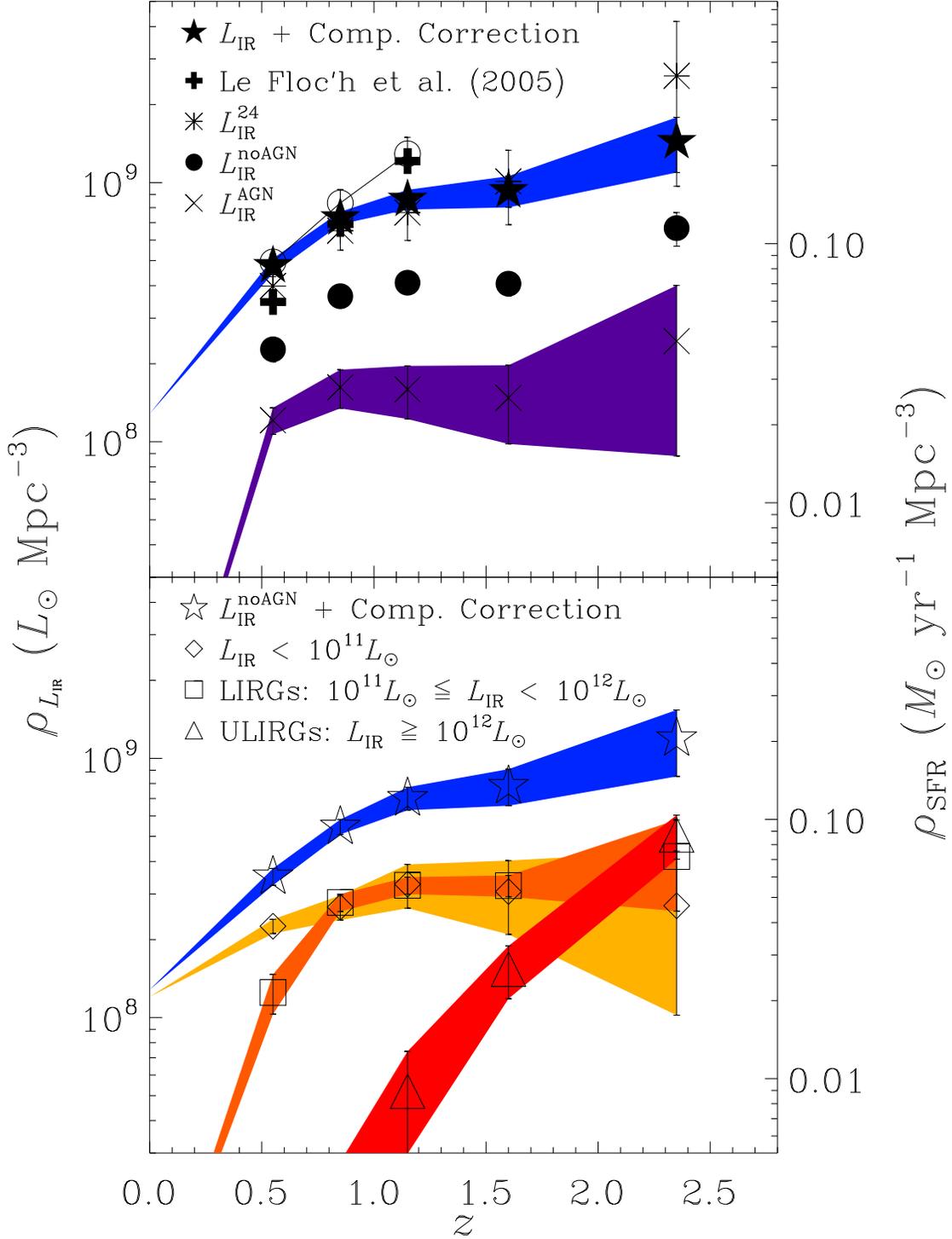}}
\caption{
The evolution of the comoving star formation and accretion-driven IR luminosity densities with redshift.  
In the top panel we plot $\rho_{L_{\rm IR}}$ calculated using 24~$\micron$ based IR luminosities (asterisks) and our best-fit IR luminosities, fit using our other photometric data, after subtracting an estimate of the AGN contribution to the total IR luminosity (filled circles).  
An estimate of the AGN luminosity density is shown as crosses, which can be considered an upper limit.  
%The total star formation driven IR luminosity density, 
The total IR luminosity density (i.e., before subtracting out an estimate of the AGN contribution), 
corrected for sources below our sensitivity limits using the luminosity function work of \citet{bm10}, is given by the filled stars in each panel (see $\S$\ref{sec-cc}).  
For comparison with other work, we plot the IR luminosity density estimates of \citet[][bold plus symbols]{el05} for our first 3 redshift bins; 
correcting our 24~$\mu$m-derived IR luminosities with their IR luminosity functions (open circles), we find that our 24~$\mu$m-derived IR luminosity density values agree, but are significantly above our best-fit SFR density determinations.   
In the bottom panel we show the total star formation driven IR luminosity density (i.e., after subtracting out an estimate of the AGN contribution), which have been corrected for sources below our sensitivity limits, as open stars.  
We also show the relative contributions to the star formation driven IR luminosity density (i.e., AGN corrected), calculated using all available photometric data, arising from low luminosity galaxies (diamonds), LIRGs (squares), and ULIRGs (triangles).  
The luminosity classi
The $z=0$ data point for the total IR luminosity density, along with contributions from normal galaxies, LIRGs, and ULIRGs, was measured by integrating the $z=0$ luminosity function of \citep{bm09}, while the $z=0$ AGN IR luminosity density was taken from \citet{xu01} and corrected for different cosmologies.  
%The filled triangle is the UV-selected ULIRG IR luminosity density taken from \citep{gm10}, which likely gives a lower limit on the IR luminosity density of ULIRGs at $z\sim3$.  .  
The shaded regions include the 1$\sigma$ uncertainties.  
\label{fig-7}}
\end{center}
\end{figure*}

Taking these improved estimates for the total IR luminosity of our 24~$\mu$m-selected sample of sources, we can investigate the evolution of the IR luminosity density $\rho_{\rm IR}$ as a function of redshift.  
We do this in five different redshift bins between $0.4 \leq z < 2.8$ (i.e., the last 5 redshift bins in Table \ref{tbl-2}).  
By using our IR luminosity estimates that have been corrected for the presence of extra infrared emission arising from embedded AGN, we convert the IR luminosity density into a SFR density $\rho_{\rm SFR}$ to examine the evolution of the star formation history of the Universe over the last $\sim$11.3~Gyr (see Figure \ref{fig-7} and Table \ref{tbl-3}).  
In this analysis we only consider 24~$\mu$m sources included in the areal coverage of the ACS imagery since these were the sources targeted for spectroscopic redshifts.  
The angular area of this region on the sky is approximately 160 arcminutes$^2$.  
A detailed discussion of the evolution of the star formation and accretion driven IR luminosity densities with redshift using our new results can be found in $\S$\ref{sec-irldev}.  
% and photometric data to calculate photometric redshifts.    
%Since our spectroscopic completeness is only $\approx$50\% among these sources, we use existing optical, NIR, and IRAC photometry to obtain photometric redshifts for a number of the sources without spectroscopic redshifts (see $\S$3.1).  

\subsubsection{Completeness Corrections for the Faint End of the Luminosity Function \label{sec-cc}}
To correct for the additional luminosity arising from galaxies having 24~$\mu$m flux densities below the 24~$\mu$m sensitivity limit (i.e., see Figure \ref{fig-6}), we employ the luminosity function derived from FIDEL data by \citet{bm09,bm10}.  
The luminosity function is parameterized by a broken power-law, whose evolution flattens once beyond $z\ga1.3$, being consistent with a luminosity evolution proportional to $(1+z)^{3.6\pm0.4}$ and $(1+z)^{1.0\pm0.9}$ below and above $z\approx1.3$, respectively.  
The corresponding density evolution for redshifts below and above $z\approx1.3$ is proportional to $(1+z)^{-0.8\pm0.6}$ and $(1+z)^{-1.1\pm1.5}$, respectively.  
At high redshifts, these corrections will depend on the assumption that the infrared luminosity function has the same logarithmic slope that is measured locally \citep[i.e.,][]{ds03}.  

By evolving the IR luminosity function appropriately for a given redshift, we integrate the luminosity contribution for galaxies having IR luminosities between 10$^{7}~L_{\sun}$ and the IR luminosity corresponding to our 24~$\mu$m upper limit of 30~$\mu$Jy at redshifts spaced by increments of $\Delta z = 0.05$ within each of the 5 redshift bins between $0.4 \leq z < 2.8$.  
%(i.e., excluding the lowest redshift bin).}  
We then averaged the completeness corrections included in each of these redshift bins, weighted by the volume of each redshift increment.  
This value was then assigned as the effective completeness correction for that redshift bin.  
The completeness corrections for these redshift bins are $\approx1.2$, 1.8, 2.9. 3.8, and $5.3\times10^{8}~L_{\sun}~{\rm Mpc}^{-3}$, respectively.  
% of that redshift bin (i.e., $z=0.4, 0.7, 1.0, 1.3, {\rm and}~1.9$).    
Similarly, we used this approach when accounting for corrections of galaxies in different luminosity classes (i.e., normal galaxies 
%- $L_{\rm IR} < 10^{11}~L_{\sun}$; 
and LIRGs).  
The values of the completeness correction for LIRGs per each redshift bin are  $\approx0.0$, 0.0, 0.058, 0.70, and $2.6\times10^{8}~L_{\sun}~{\rm Mpc}^{-3}$, respectively.  
The difference between the total completeness correction and these values correspond to the completeness correction for normal galaxies alone.  

\begin{figure}
\plotone{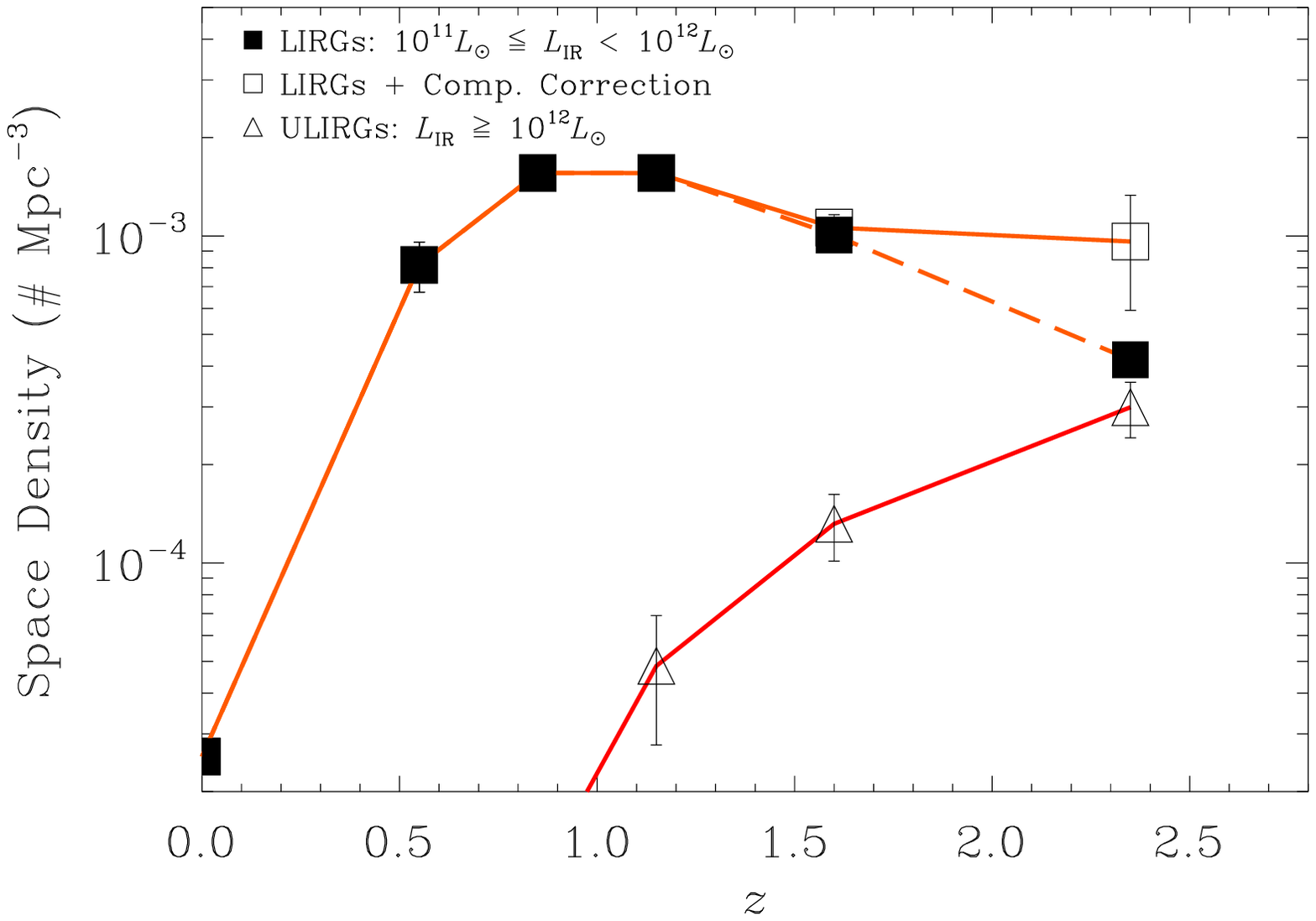}
\caption{
The space density evolution of 24~$\mu$m-selected LIRGs (squares), and ULIRGs (triangles).  
The contribution from LIRGs before applying a completeness correction (see $\S$\ref{sec-cc}) are given by the filled squares.  
Values at $z=0$, along with the first two redshift bins for which we do not detect ULIRGs (i.e., $z=0.55$ and 0.85) were taken from \citet{bm09}.  
\label{fig-8}}
\end{figure}

In Figure \ref{fig-8} we plot the space density evolution for LIRGs and ULIRGs.  
Also shown is the completeness corrected number density for LIRGs using the technique described above.  
These corrections are $\approx0.0$, 0.0, 0.054, 0.54, and $1.1\times10^{-3}~{\rm Mpc}^{-3}$, respectively.  
%Similarly, we correct for incompleteness among the LIRGs in each redshift bin using the same luminosity functions resulting in corrections of $\approx0.0$, 0.0, 0.054, 0.54, and $1.1\times10^{-3}~{\rm Mpc}^{-3}$, respectively.  

\subsubsection{Propagations of Errors for IR Luminosity Densities}
There are four primary sources of uncertainty in estimating the IR luminosity densities: photometric redshifts, photometric uncertainties, bolometric corrections, and completeness corrections.  
%We assign uncertainties to the IR luminosity densities plotted in Figure \ref{fig-7} based on the uncertainties in the SED fitting.  
These are derived using a Monte Carlo technique from both photometry, and where applicable, the photometric redshift.  % (see $\S$3.1 and $\S$3.2.1).   
%the photometric and, in the case of those sources for which we only have a photometric redshift,  redshift uncertainties (see $\S$3.2.1).    
The additional uncertainty of the scatter in our empirical correction for the overestimation of the 24~$\mu$m-derived IR luminosities, which is  $\approx$27\% for sources with $L_{\rm IR}^{24} < 10^{12}~L_{\sun}$ and $\approx$42\% about the fitted regression line for sources with $L_{\rm IR}^{24} \geq 10^{12}~L_{\sun}$ (see the right panel of Figure \ref{fig-5}) is then added to the fitting errors in quadrature for each galaxy.      
Similarly, we include an uncertainty for the AGN subtraction based on the residual dispersion (i.e., $\approx$66\%) in the empirical correlation \citep[i.e., right panel of Figure 8 in][]{ejm09a} used to derive Equation \ref{eq-agnfrac}.  
This uncertainty is included in the quadrature sum above.  

%The uncertainty in the luminosity completeness corrections are based on the dispersion in the bolometric corrections (observed-frame $L_{\rm IR}/\nu L_{\nu} (24~\mu$m) ratios at each redshift bin in Table \ref{tbl-2}) for the \citet{ce01} templates having effective dust temperatures of $36\pm7$~K, where a dust grain emissivity  $\beta_{\rm dust} = 1.6$ was assumed; this temperature range has been estimated from average galaxy SEDs at these redshifts \citep{dye07}.  
%The corresponding bolometric correction uncertainties are 8, 16, 21, 32, and 62\% for each redshift bin, respectively.  
The uncertainty in the completeness corrections for the faint end of the luminosity functions are based on the dispersion in the bolometric corrections (observed-frame $L_{\rm IR}/\nu L_{\nu} (24~\mu$m) ratios at each redshift bin between $0.4 \leq z < 2.8$ in Table \ref{tbl-2}) for the \citet{ce01} templates.  
Assuming effective dust temperatures of $T_{\rm dust} = 36\pm7$~K, and a dust grain emissivity of $\beta_{\rm dust} = 1.6$ \citep[e.g., this temperature range has been estimated from average galaxy SEDs at these redshifts;][]{dye07}, we find that the corresponding bolometric correction uncertainties are 8, 16, 21, 32, and 62\% for each redshift bin, respectively.  
These uncertainties are then added in quadrature to the uncertainties on the total IR luminosity density of each redshift bin, which was based on the propagation of the uncertainties on the IR luminosities of each source included in that redshift bin.  
The shaded regions plotted in Figure \ref{fig-7} illustrate the 1-$\sigma$ uncertainties.

\begin{figure}
\plotone{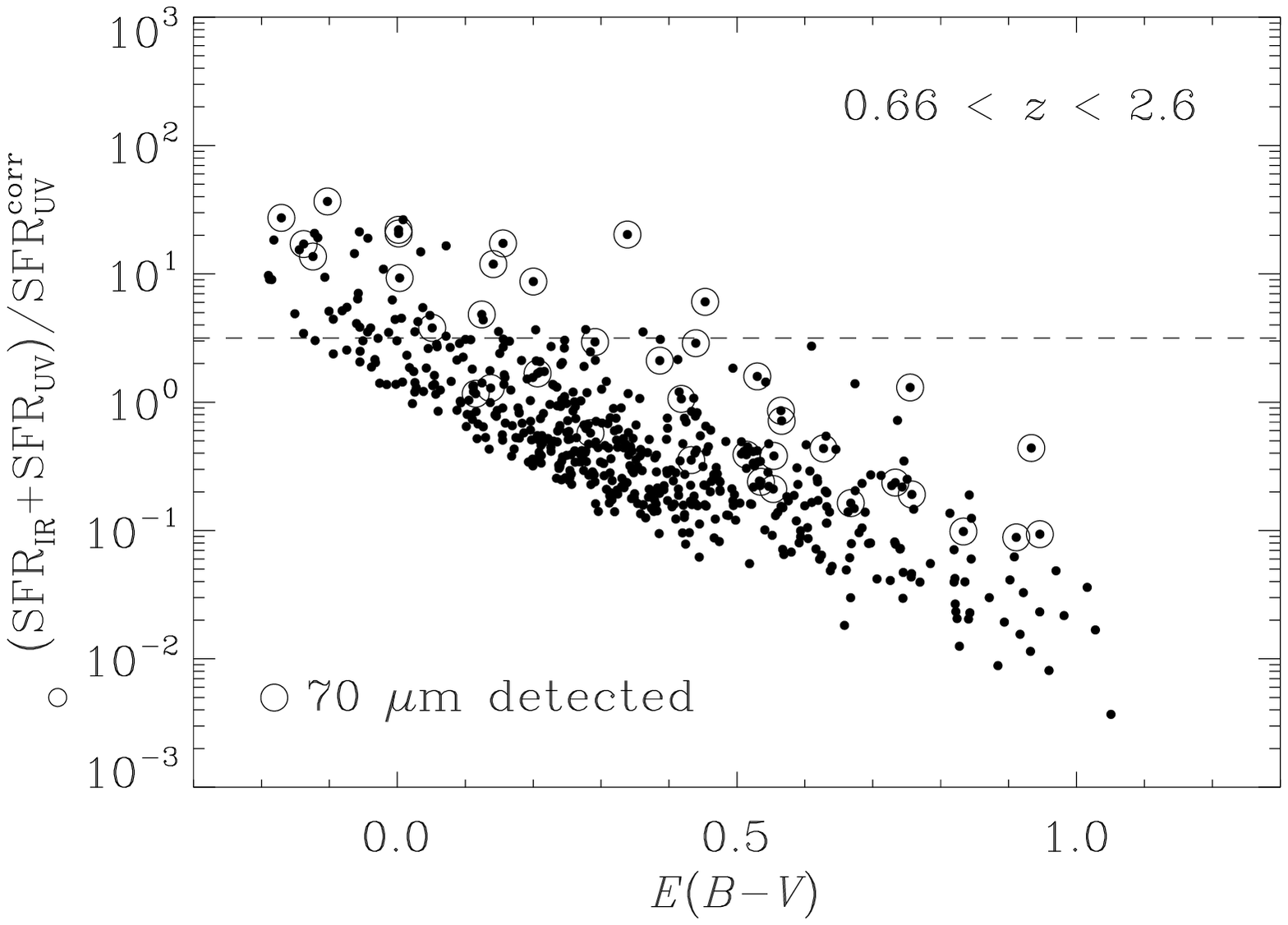}
%\plottwo{f11c_A2.eps}{f11d_A2.eps}
%\plottwo{f5e.eps}{f5f.eps}
\caption{ 
We plot the ratio of IR derived SFRs estimated from SED fitting (plus UV derived SFRs uncorrected for extinction) to extinction corrected UV SFRs against the color excess $E(\bv)$ inferred from fitting the rest-frame UV ($1250 - 2600$~\AA) slope using $UBViz$ optical imaging data; those galaxies which are 70~$\micron$ detected are identified with circles.    
IR luminosities have not been corrected for AGN contributions, which makes little difference for the observed trend.   
%In the top two panels we plot only those galaxies lying within a redshift range between $1.4 < z < 2.6$ (i.e., approximately the $BzK$ selection range).    
%This is done to illustrate the difference in extinctions derived by using color relations (top left panel; i.e., \citet{ed04}) to that when a proper slope is fit (top right panel).  
%The plus symbols in the top right panel were generated using our revised extinction corrections (see $\S$6).  
%In the bottom panels this is done for all galaxies having spectroscopic redshifts for which fitting the rest-frame UV continuum was possible; the minimum redshift among these galaxies is $z \approx 0.66$.  
%In the bottom left panel, IR SFRs were derived from SED fitting 24~$\micron$ flux densities alone while, for the bottom right panel, IR based SFRs were derived from SED fitting  the 16, 24, and 70~$\micron$ photometry; those galaxies which are 70~$\micron$ detected are identified with circles.  
\label{fig-9}}
\end{figure}

\begin{figure*}
\plottwo{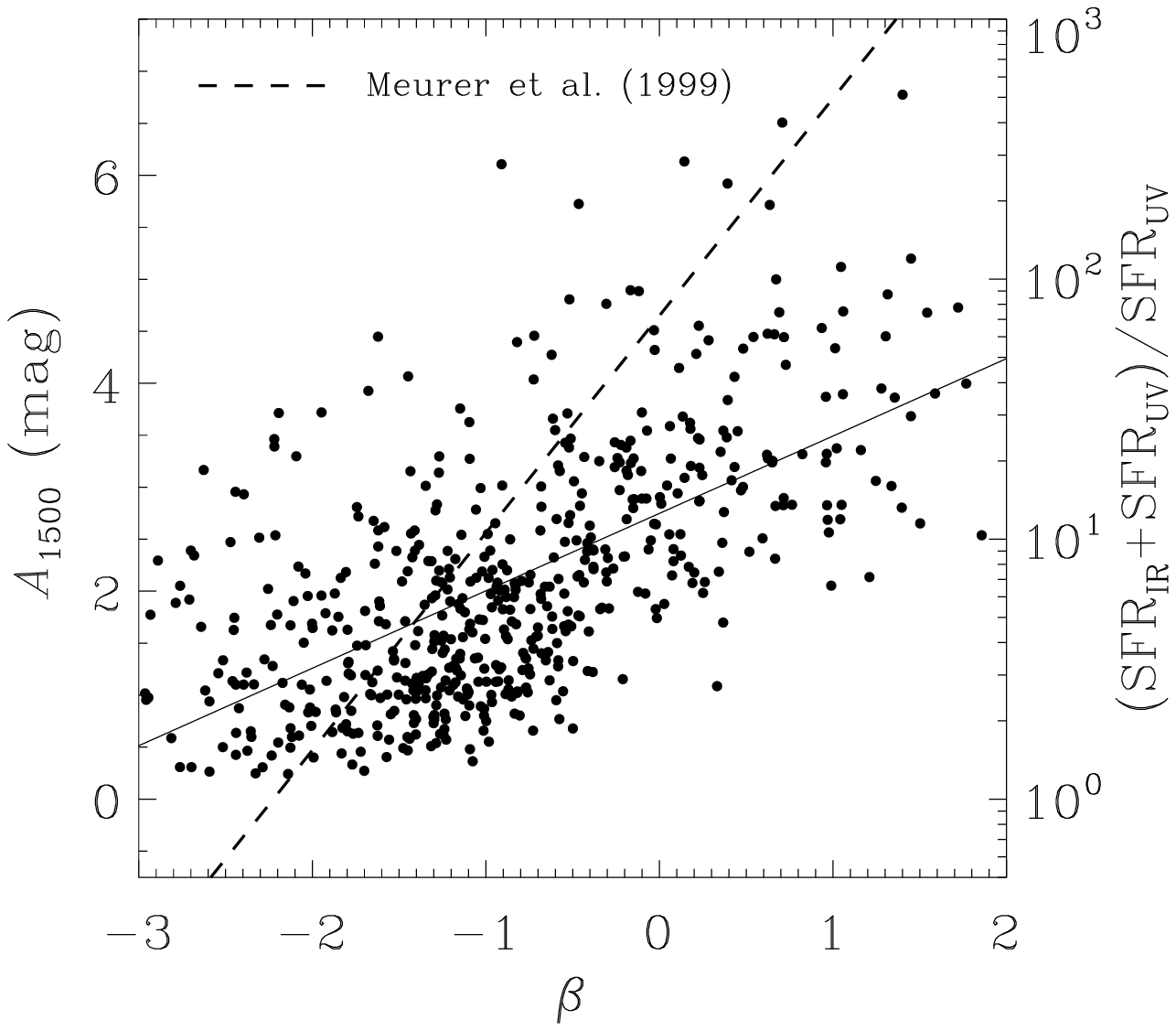}{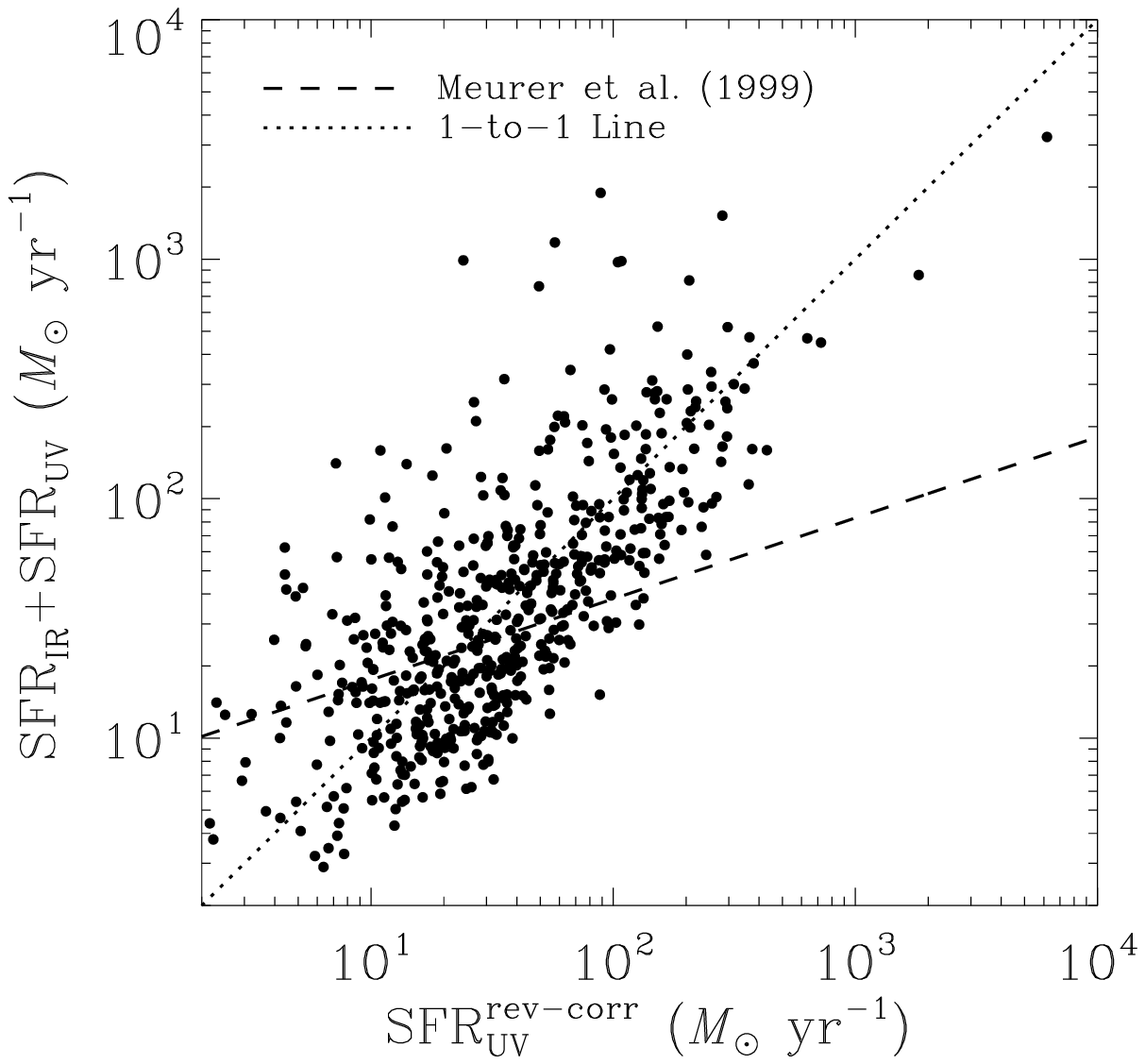}
\caption{
In the left panel we plot our estimate of $A_{1500}$, which is derived from the ratio of the total (IR $+$ observed UV) to the observed UV SFR, against the measured UV spectral slope, $\beta$, for those galaxies shown in Figure \ref{fig-9}.     
As in Figure \ref{fig-9}, IR luminosities have not been corrected for AGN contributions.  
The over plotted solid-line is the ordinary least squares fit to  $A_{1500}$ versus $\beta$ (Equation \ref{eq-irxb}), while the dashed-line indicates the UV extinction correction of \citet[][i.e., Equation \ref{eq-extcor}]{gm99}.     
In the right panel we plot the total (IR $+$ observed UV) SFR against our revised extinction corrected UV SFRs using the fit from the left panel.  
The dotted-line is a one-to-one line.  
The dashed-line indicates the fit to the IR $+$ observed UV SFR versus the UV corrected SFR when using the UV spectral slope to correct for reddening given by Equation \ref{eq-extcor}.  %}\citet[][i.e., Equation \ref{eq-extcor}]{gm99}.  
This discrepancy clearly shows that the UV extinction correction using \citet{gm99} is not appropriate for this sample of 24~$\mu$m-selected sources.  
\label{fig-10}}
\end{figure*}

\section{Comparison Between IR and Extinction-Corrected UV SFR\MakeLowercase{s} for 24~$\micron$-Selected Sources}  
%Revised UV Extinction Corrections}
%\noindent{\bf KEEP DISCUSSION ABOUT EXTINCTION CORRECTIONS?}\\
Having our improved estimates for IR luminosities, we now look to see how well local extinction corrections, based on the UV spectral slope, work for our sample of 24~$\mu$m-selected galaxies.  % using our improved estimates for IR luminosities.  
A full treatment investigating local extinction correction among 24~$\mu$m-selected galaxies in the $BzK$ redshift range can be found in Appendix C.  

\subsection{Typical Correction for UV Extinction}
UV-based SFRs (SFR$_{\rm UV}$) are calculated using the rest-frame 1500~\AA~ specific luminosities for each source using the $UBViz$ photometry and the conversion given in \citet{rck98} as described in \citet{ejm09a}.    
We estimate the amount of extinction at 1500~\AA~ using the empirical relation between the UV spectral slope spanning $1500-2600$~\AA, defined as $\beta$, to the total amount of extinction at 1600~\AA~\citep{gm99}, along with modeled extinction curves \citep[i.e.][]{wd01,btd03}, such that, 
%between the amount of extinction at 1600~\AA~ and the slope of the UV continuum between $1250-2600$~\AA~, defined as $\beta$, leading to the relation  
\begin{equation}
\label{eq-extcor}
\left(\frac{A_{1500}}{\rm mag}\right)= 4.65 + 2.09 \beta.
\end{equation}
Since the redshifts of the sample galaxies are at $z < 3$, any correction to the UV slope due to absorption by the intergalactic medium is likely negligible.   
Sources having $\beta < -3$ or $\beta > 2$ are excluded in this analysis as this is outside the range for which the UV spectra slope has been calibrated to extinction.  
These sources were generally those having (low) redshifts for which $\beta$ was determined using two data points that included the ground-based $U$-band data, which is less sensitive and has much coarser resolution than the ACS imaging.     
The minimum redshift for which we have calculated a reliable extinction correction UV luminosity is $z \approx 0.66$.  
Extinction corrected UV SFRs are denoted as ${\rm SFR_{\rm UV}^{\rm corr}}$.  

\subsection{Applicability to 24~$\mu$m Detected Sources}
%This is done for the entire sample, as well as the subset of galaxies lying in the redshift range between $1.4 < z <2.6$ (i.e., approximately the $BzK$ selection range).  
%\subsection{Among $z<3$ Sources}
%In the bottom left panel of 
In Figure \ref{fig-9} we plot the ratio of $\rm IR+UV$ to UV corrected SFRs
%, where the IR SFR was calculated using the 24~$\micron$ derived IR luminosity, 
versus the estimated $E(\bv)$ color excesses (i.e., $E(\bv) = 0.123 A_{1500}$).    
For these comparisons with the UV SFRs, IR luminosities have not been corrected for the presence of AGN, however using the AGN subtracted values does not significantly affect the results.  
We note that the minimum redshift among these sources, where we could obtain a reliable estimate of the UV spectral slope, is $z \approx 0.66$.  
We have also limited this comparison to sources with $z < 2.6$.    
%Similarly, in the bottom right panel of Figure \ref{fig-11}, we show the same ratio of SFRs, this time using our best-fit IR luminosities to calculate the IR based SFR, against the corresponding extinction estimates.  
%In both plots 
The UV corrected SFRs appear to be overestimated by more than a factor of $\ga$2, on average, among the 24~$\mu$m-selected sources plotted, and we see a clear trend of decreasing ratio of IR$+$UV to UV corrected SFRs with increasing extinction suggesting that the UV slope is overestimating the extinction in a large number of sources.
%Among all the sourced plotted, the UV corrected SFRs appear to overestimate the actual SFR by more than a factor of $\approx$2 due to an overcorrection for extinction using Equation \ref{eq-extcor}.  % by a factor of more than $\approx$2
%We also see that the dispersion about these trends appears to be larger (i.e., a factor of $\sim$3.6) when using the 24~$\micron$ derived IR luminosities compared to when using our best-fit IR luminosities to derive IR SFRs; the decreased dispersion is especially evident towards the lower extinction values.  

The fact that there is a trend at all suggests that the treatment for extinction correcting the UV flux densities is not adequate for our sample.  
To quantify this, we plot an estimate of $A_{1500}$, which is derived from the ratio of the total (IR $+$ observed UV) to the observed UV SFR, against the UV spectral slope, $\beta$ (see the left panel of Figure \ref{fig-10}).  
A clear trend of increasing extinction with $\beta$ is found, for which an ordinary least squares fit results in the relation
\begin{equation}
\label{eq-irxb}
\left(\frac{A_{1500}}{\rm mag}\right) = 2.75\pm0.06 + (0.75\pm0.04)\beta.
\end{equation}
Our sample of 24~$\mu$m-selected galaxies typically fall well below the UV extinction correction of \citet[][i.e., Equation \ref{eq-extcor}]{gm99}, which is plotted as a dashed line.   

Taking these revised estimates for the UV extinction, we correct our UV luminosity and derive revised UV corrected SFRs ($\rm SFR_{\rm UV}^{rev-corr}$).  
In the right panel of Figure \ref{fig-10} we plot the IR$+$(observed) UV SFRs against the revised, extinction corrected UV SFRs which exhibits a general one-to-one trend, albeit with a significant amount of (factor of $\sim$3) scatter.
%For comparison, the sources in the top right panel of Figure \ref{fig-11} are re-plotted using the revised extinction corrections as plus symbols.    
Also shown in the right panel of Figure \ref{fig-10} is the result from fitting the IR$+$(observed) UV SFRs with the extinction corrected UV SFRs using the relation taken from \citet{gm99}.  
A clear discrepancy is found between these two fits suggesting that the extinction correction prescription given by \citet{gm99} is not appropriate for our sample of 24~$\mu$m-selected sources.  
Extinction corrected UV luminosities are a factor $\approx$2 larger using the local (starburst) relation of \citet[][i.e., Equation \ref{eq-extcor}]{gm99} compared to our revised values  using Equation \ref{eq-irxb}.    

We believe this result may arise in part due to the age-extinction degeneracy among the lower redshift (i.e., $0.66 \la z \la 1.4$) objects in our sample.  
Attenuation relations flatter than that of  \citet{gm99} have also been reported for samples of `normal' star-forming galaxies \citep[e.g.,][]{efb02,kcb04,vb05}, 250~$\mu$m-selected galaxies at $z<1$ \citep{vb10}, as well as UV-selected $z\sim1$ galaxies when comparing to stacked X-ray SFRs \citep{esl05}.   
If the extinction is relatively lower in these objects, the UV slope will no longer be as sensitive to the amount of extinction due to the importance of variations arising from contributions by a galaxy's old stellar population.  
Another explanation for a flatter attenuation relation could be due to a steeper UV extinction curve among our sample of 24~$\mu$m-selected sources.  
While most normal $z\sim2$ LBGs (i.e., UV-selected starbursts with $10^{10}~L_{\sun} \la L_{\rm IR} \la 10^{12}~L_{\sun}$) follow the \citet{gm99} relation, young ($\la$100~Myr) LBGs tend to fall below the local starburst relation which has been attributed to a steeper, more SMC-like, UV extinction law in these sources \citep{nr06,nr10}.  
This physical effect may also play a role in placing our sample of 24~$\mu$m-selected galaxies below the \citet{gm99} relation in the left panel of Figure \ref{fig-10}.

%\begin{figure}
%\begin{center}
%\scalebox{1.15}{
%\plotone{f10.eps}}
%\caption{
%We plot $q_{\rm IR}$ versus redshift for all 70~$\mu$m detected sources (open circles).  
%For galaxies detected at 70~$\micron$ but not at 20~cm, we indicate that the IR/radio ratio is a lower limit by plotting an upward arrow.  
%The over plotted dashed lines indicate the expected evolution in the FIR-radio correlation due to increased IC losses of CR electrons off of the CMB at increasing redshift for galaxies having different intrinsic magnetic field strengths \citep{ejm09b} and a constant radiation field energy density calculated using the median radio size for the 22 mid-infrared spectroscopic sample. 
%Unlike the evolutionary tracks included in \citet{ejm09b}, which only considered competition between synchrotron losses and IC scattering of CR electrons off of the CMB, these tracks also include additional energy loss terms arising from ionization and bremsstrahlung processes, as well as escape.  
%The bold solid line corresponds to the evolutionary track by taking the median minimum energy magnetic field and radiation field strengths among the 70~$\mu$m detected sources.  
%Clearly, the sources do not appear to be following these tracks.  
%\label{fig-13}}
%\end{center}
%\end{figure}

\section{Discussion}
\label{sec-irldev}
Following the results of \citet{ejm09a}, we use deep imaging at 70~$\micron$ of GOODS-N, as part of the FIDEL survey, to calculate the IR luminosities for 24~$\micron$ detected sources.  
Using our improved bolometric corrections and estimates for the contribution of AGN (see $\S3.1$), we characterize the evolution of the star formation and AGN powered IR luminosity density out to $z \sim2.8$.  
% having spectroscopic redshifts to see how the additional 70~$\micron$ data helps to better constrain our SED fitting.  
%For sources fainter than $\sim$200~$\mu$Jy, the additional 70~$\micron$ data does not help to better constrain the SED fitting since only only upper-limits can be determined.  
%Results associated with these objects are therefore not included in this following discussion.  

\subsection{Uniqueness of Sample and Analysis}
Here we highlight the major differences between our work for deriving the co-moving SFR and accretion energetics over the past $\sim$11 Gyr compared to others in the literature.  
First, we are working with a complete sample of 24~$\mu$m-selected sources derived from the deepest 24~$\mu$m imaging carried out by {\it Spitzer}.  
The fact that these observations were taken for the GOODS-N field provides a treasury of optical and NIR ancillary data for assigning redshifts (spectroscopic and photometric) to $\approx$94\% or the sample.  
Additionally, deep 70~$\mu$m, and 850~$\mu$m data were available for a subset of the sample, allowing us to estimate the true IR luminosities of each source based on the deep 24~$\mu$m photometry alone (i.e., without having to rely on stacking analyses).  
Deep mid-infrared spectroscopy has allowed us to empirically estimate the fractional AGN contribution to each IR luminosity, resulting in the first measure of obscured AGN energetics with redshift, in addition to measuring contributions to the SFR density as a function of redshift for LIRGs and ULIRGs.  
Consequently, we have derived a new UV extinction correction based on the measure of the UV spectral slope for 24~$\mu$m-selected sources using these new, more accurate, estimates for IR luminosity.  
%\begin{itemize}
%\item Complete 24~$\mu$m sample in GOODS-N w/ redshift (phot + spec) completeness of 95\%.
%\item Deep MIR spectra, 70, and 850~$\mu$m data to accurately calibrate true $L_{\rm IR}$ estimates based on 24~$\mu$m photometry alone.
%\item The first measure of obscured AGN energetics with redshift in addition to LIRGs and ULIRGs
%\item New derivation of UV extinction corrections for 24~$\mu$m-selected sources using new, more accurate $L_{\rm IR}$s.  
%\end{itemize}

\subsection{An Accounting of the Star Formation and Accretion Histories to $z \sim 3$}
In the top panel of Figure \ref{fig-7} we plot the IR luminosity density versus redshift for galaxies included in the last 5 redshift bins of the bottom panel of Table \ref{tbl-2} (i.e., 24~$\mu$m-selected galaxies covered in the ACS imaging).  
For the directly detected 24~$\mu$m sources, we plot independently the contribution from star formation (solid circles) and accretion processes (crosses) to our best-fit IR luminosities.  
We also plot the IR luminosity density evolution one would predict based on using the 24~$\mu$m-derived IR luminosities (asterisks).  
%We plot 
The corresponding IR luminosity density estimates for our first three redshift bins taken from \citet{el05}  are also given (bold plus symbols); 
after applying a completeness correction based on their luminosity function work (open circles), we find that our 24~$\mu$m-derived IR luminosity density estimates are consistent.  
However, when the best-fit IR SFR is used (filled stars), we find that the \citet{el05} results are significantly above our values.  
This is due to the more precise bolometric correction and AGN subtraction used in this paper.  

%We estimate the total {\it star formation driven} IR luminosity density evolution by including a completeness correction to our AGN corrected best-fit IR luminosities (filled stars; see $\S$3.2.1); 
We estimate the total and {\it star formation driven} IR luminosity density evolution by including a completeness correction to our best-fit IR luminosities before (top panel; filled stars) and after (bottom panel; open stars) correcting for AGN (see $\S$3.2.1); 
even with the completeness correction, the AGN-corrected values are still well below what the 24~$\mu$m-derived IR luminosity density suggests.  
By using a more robust means of excising out the AGN contribution to the total IR luminosities, our value of the SFR density at $z\approx2.35$ is $\rho_{\rm SFR}  \approx 0.21\pm0.06~M_{\sun}~{\rm~yr^{-1}~Mpc^{-3}}$.  
%This value is slightly below (i.e., $\approx$15\%) that reported by \citet{nr08} for a very similar redshift bin (i.e., $1.9 \leq z < 2.7)$, however we note that these authors did not account for a contribution from AGN.  
While this value is consistent with that reported by \citet{nr08} for a very similar redshift bin (i.e., $1.9 \leq z < 2.7)$, these authors did not account for a contribution from AGN.  
The fractional contribution of AGN to the total IR luminosity density is found to decrease from $\la$25\% to $\la$15\% with increasing redshift between $0.55 \la z \la 2.35$.  
%Accordingly, the total IR luminosity density in our highest redshift bin is only $\approx$5\% larger than that of \citet{nr08}.  
On the other hand, the UV-corrected SFR density reported by  \citet{nr08}, which assumes an average extinction correction of a factor of 4.5, is $\approx$12\% larger than what is derived here.  
%, although slightly below (i.e., $\sim$80\% of), the value reported by \citet{nr08}.  

By averaging the SFR densities in our highest two redshift bins, we obtain a $z\sim2$ SFR density of $\rho_{\rm SFR}\sim0.17\pm 0.06~M_{\sun}~{\rm~yr^{-1}~Mpc^{-3}}$ which we can compare to other $z\sim2$ values quoted in the literature.  
This value is slightly larger (i.e., $\approx$15\%) than the $z\sim2$ census value of $0.15\pm0.03~M_{\sun}~{\rm~yr^{-1}~Mpc^{-3}}$, computed from directly detected (optical, NIR, and submillimeter) galaxies \citep{nr05}, which can be considered a lower limit for the SFR density at this epoch.    
This value is also $\approx$50\% larger than the $z\sim2$ SFR density reported by \citet{kc07}, which is most likely due to their using a much flatter faint-end slope for their infrared luminosity function as pointed out by  a number of authors \citep[e.g.,][]{nr08,gr10,bm10}.

%\subsubsection{Decomposing the IR Luminosity Density versus Redshift}
In the bottom panel of Figure \ref{fig-7} we decompose the star formation driven IR luminosity densities into contributions from normal galaxies (diamonds), LIRGs (squares), and ULIRGs (triangles).  
We note that sources are first separated into IR luminosity classes (normal galaxies, LIRGs, and ULIRGs) based on their total IR luminosity (i.e., the IR luminosity before subtracting out an AGN contribution).    
By $z\approx0.85$ there is an equal contribution to the SFR density from normal galaxies and LIRGs.  
The LIRGs 
%then peak and 
continue to contribute equally with the population of normal galaxies until $z\approx1.6$, at which point the contribution of normal galaxies begins to decline out to our highest redshift bin at $z\approx2.35$.  % which LIRGs contribution remains roughly constant.  
We note that the completeness correction, which is unconstrained from the data presented here, is largest for normal galaxies in our highest redshift bin, reflected in the large uncertainties.    
The trend found for the space density evolution of LIRGs (Figure \ref{fig-8}) is slightly different than their luminosity density evolution.    
The space density of LIRGs appears to peak between $0.85\la z \la1.15$ ($\approx1.6\pm0.1\times10^{-3}~{\rm Mpc^{-3}}$) and remains flat until beyond $z\ga1.15$ where it decreases to $\approx1.0\pm0.4\times10^{-3}~{\rm Mpc^{-3}}$ at $z\approx2.35$.  

The ULIRGs, however, do not appear to account for more than $\sim$20\% of the SFR density until redshifts of $z\ga$1.6, at which point they become roughly comparable with the contribution from LIRGs 
%the total IR luminosity density 
once beyond $z\ga2$, contributing at the $\approx$42\% level by $z \approx 2.35$ (see Table \ref{tbl-3}).    
%On the other hand, 
Our result is larger than the 30\% contribution of ULIRGs to the total SFR density reported by \citet{nr08} in this redshift range (i.e., $1.9 \la z \la 2.8$).    
On the other hand, the total contribution from LIRGs (37\%) and ULIRGs (33\%) at $z\sim2$ is $\approx$70\%, significantly smaller than the $\approx$93\% contribution from LIRGs (48\%) and ULIRGs (45\%) reported by \citet{gr10}.  
Similarly, our result is in contrast with claims that ULIRGs alone dominate the SFR density at $z\sim2$ \citep[e.g.,][]{pg05}.  
We attribute these differences between previous studies to our improved IR luminosity estimates and accounting for AGN, which preferentially affects the most luminous 24~$\mu$m sources.     
In the case of \citet{gr10}, these authors removed type-1 quasars in their SFR density calculations, assuming the luminosities of such sources are entirely accretion driven, but did not include a contribution from the bulk of the population as done here; 
they note that the removal of these sources did not have a large impact on their reported SFR densities, and is thus not the reason for differences between our results.  
%, and slightly smaller than equally split 90\% contribution of LIRGs and ULIRGs to the SFR density at these epochs reported by \citet{pg05,gr10}.       

Similar to \citet{bm10}, we find that the IR luminosity density of LIRGs remains roughly flat between $1.15 \la z \la 2.35$, however we also find the contribution from ULIRGs, and the total SFR density, may increase with redshift over this range.    
These authors also find that ULIRGs do not dominate the SFR density at $z\sim2$, albeit they report an even smaller contribution from ULIRGs (i.e, 17\%) than the 33\% reported here.  
While ULIRGs appear to contribute significantly to the SFR density at $z\approx2.35$, we note that the contribution made by ULIRGs to the total SFR density of the Universe integrated between $0 \la z \la 2.35$ is small, being $\approx$20\%, where as LIRGs and normal galaxies contribute equally at the $\approx$40\% level.    
Furthermore, as shown in Figure \ref{fig-8}, the space density of ULIRGs is significantly less (i.e., a factor of $\approx$3) than LIRGs at $z\approx2.35$.  
%This finding is in contrast with that of \citet{bm10}, who find that the IR luminosity density is constant between $z\approx 1.3$ and 2.3.  

While the contribution of ULIRGs appears to be large in our final redshift bin, it is not clear that this trend should continue to higher redshifts.  
%In the bottom panel of Figure \ref{fig-7} we show the $z\sim3$ UV-selected ULIRG IR luminosity density taken from \citep[filled triangle;][]{gm10}.  
Recent work by \citet{gm10} report on the IR luminosity density of UV-selected ULIRGs at $z \sim 3$, being $\sim7.5^{+6.5}_{-4.5}\times 10^{7}~L_{\sun}~{\rm Mpc}^{-3}$.  
This number is significantly smaller than (i.e., $\approx$15\% of) our inferred SFR density for $z\approx2.35$ ULIRGs, and provides a lower limit on the fractional contribution of ULIRGs to the SFR density at $z\sim3$ since the UV selection is likely to miss more obscured, UV-faint ULIRGs.    
%suggesting a steep decline in the fractional contribution of ULIRGs to the SFR density.  
We also note that the space density of the 24~$\mu$m-selected ULIRGs at $z\approx2.35$ (i.e., $\approx3.0\pm0.6\times10^{-4}~{\rm Mpc^{-3}}$) is a factor of $\approx$20 times larger than the space density of $z\sim3$ UV-selected ULIRGs reported by \citet{gm10}.    

It is possible that our final redshift bin may be over predicting the SFR density due to an under estimation of AGN at these epochs since we are assigning the AGN contribution based on a relation which was derived for a heterogenous sample of only $\approx$20 galaxies \citep{ejm09a}.  
However, even by assuming that the IR luminosities estimated for all hard X-ray detected sources, along with those sources having an IRAC SED described by a power law, are completely AGN powered, this does not flatten the SFR density evolution nor bring the ULIRG contribution below that of the normal galaxies in our highest redshift bin; the SFR density for the final redshift bin decreases to $\rho_{\rm SFR} \approx 0.19~M_{\sun}~{\rm~yr^{-1}~Mpc^{-3}}$ while the fractional contribution by ULIRGs decreases to $\approx$39\%.  

While our results suggest an increase to the IR luminosity density out to z$\approx$2.35 and that ULIRGs may contribute significantly to the SFR density beyond $z\ga2$, we note that the uncertainties in our final redshift bin are quite large.  
%Finally, we note that, 
%Within uncertainties (i.e., $2\sigma$), our results are consistent with the scenario in which the total SFR density remains constant between $1.15 \la z \la 2.35$, and normal galaxies, LIRGs, and ULIRGs contribute to the total IR luminosity density at similar levels.    
Deep surveys with {\it Herschel}, such as the GOODS-{\it Herschel} open time key project (PI: D. Elbaz) will likely settle this ambiguity by properly measuring the peak of the FIR SEDs for LIRGs and ULIRGs out to redshifts of $z\sim2$ and $z\sim4$, respectively.

\section{Conclusions}
In the present study we have built on the findings of \citet{ejm09a} to determine how the star formation and accretion driven IR luminosity densities have evolved over the last $\approx$11.3 Gyr.  
This has been made possible through improved estimates of IR luminosities, achievable by further constraining fits to local SED templates using deep MIPS imaging at 70~$\micron$ taken as part of FIDEL.  
%affects current pictures on the occurrences of IR excess sources, as defined by \citet{ed07a,ed07b}, and how the FIR-radio correlation and IR luminosity density evolves with redshift.    
Our conclusions can be summarized as follows:

\begin{enumerate}
\item
IR ($8-1000~\micron$) luminosities derived by SED template fitting using observed 24~$\micron$ flux densities alone overestimate the IR luminosity by a factor of $\sim$4 when their 24~$\mu$m-derived IR luminosity is $\ga3\times10^{12}~L_{\sun}$.  
% compared to when additional 16 and 70~$\micron$ detections (and/or upper limits) are included in the fits.  
%We find a similar discrepancy, although less dramatic (i.e., an overestimation by a factor of $\sim$1.6) for galaxies in redshift range between $1.9 < z < 2.8$.   
This discrepancy appears to be the result of high luminosity sources at $z\gg0$ having far- to mid-infrared ratios, as well as aromatic feature equivalent widths, which are more typical of lower luminosity galaxies in the local Universe.  
%This result suggests that local SED templates do not properly characterize the contribution from aromatics for high luminosity galaxies at increasing redshifts as these galaxies require cooler templates than local galaxies of similar luminosity.   

%\item
%Using simple color relations to correct for UV extinction can affect the associated UV derived SFRs.  
%For example, using the color relation presented by \citet{ed04} for galaxies in a redshift range between 1.5 and 3 results in a median SFR which is $\sim$0.65 times that when a proper fit is made to the UV slope.

\item
%The star formation driven IR luminosity density appears to increase out to $z\sim2.35$ which the AGN driven IR luminosity density appears to remain relatively flat beyond a redshift of $z\ga1$.  %is found to peak at $z\sim0.85$; 
After accounting for star formation and AGN contributions to the total IR luminosities, we find that AGN and star formation activity appear to roughly track one another with AGN typically accounting for $\la$18\% of the total IR luminosity density integrated between $0\la z \la 2.35$.  
The AGN fraction slightly decreases from $\la$25\% to $\la$15\% with increasing redshift between $0.55 \la z \la 2.35$.  
Our observations also hint that the star formation driven IR luminosity density may increase with redshift between $1.15 \la z \la 2.35$, however, within uncertainties, our results are consistent with a flat evolution over this redshift range.    
%We also find that, within errors, the star formation driven IR luminosity density is consistent with being flat between $1.15 \la z \la 2.35$.  

\item
The SFR density (i.e., IR luminosity density corrected for AGN contamination) is dominated by normal galaxies and LIRGs at comparable levels (i.e., each at $\approx40-50$\%) between $0.85 \la z \la 1.6$.  
%exhibiting relatively constant contributions over this redshift range.  
LIRGs continue to contribute at a similar level out to $z\approx 2.35$ which is in contrast with the ULIRGs, which transition to becoming a significant contributor to the SFR density (i.e., comparable with LIRGs) only once beyond $z \ga2$.  
%$z\sim1.15$, whose contribution remains relatively constant out to $z\sim2.35$.  
%This is in contrast with the ULIRGs which transition to becoming a significant contributor to the SFR density beyond $z \ga2$.  

\item
Among our sample of 24~$\mu$m-selected sources, we find that local prescriptions used to estimate UV extinction corrections based on the UV spectral slope typically overestimate the true extinction by a factor of $\ga$2.  
Accordingly, using our improved estimates for IR luminosity, we have derived a new UV extinction correction based on the UV spectral slope which may be more appropriate for 24~$\mu$m-selected sources.  

\end{enumerate}

\acknowledgments
We thank members of the GOODS team who contributed to the data reduction and photometric catalogs for the various data sets used here, particularly Norman Grogin, Yicheng Guo, Joshua Lee, Kyoungsoo Lee, and Harry Ferguson.  
We also thank G.~Morrison for useful discussions.  
%We would also like to thank the anonymous referee for useful suggestions which helped to improve the quality of this paper.  
This work is based on observations made with the Spitzer Space Telescope, which is operated by the Jet Propulsion Laboratory, California Institute of Technology, under a contract with NASA. Support for this work was provided by NASA through an award issued by JPL/Caltech.
Partly based on observations obtained with WIRCam, a joint project of CFHT, Taiwan, Korea, Canada, France, at the Canada-France-Hawaii Telescope (CFHT) which is operated by the National Research Council (NRC) of Canada, the Institute National des Sciences de l'Univers of the Centre National de la Recherche Scientifique of France, and the University of Hawaii.  
%\clearpage

\appendix

\section{IR Luminosity and SFR Comparisons}
%\section{Results}
\citet{ejm09a} compared IR luminosities derived from 24~$\mu$m data alone to those derived when additional measurements are available, such as 16 and 70~$\mu$m photometry, submillimeter data, and mid-infrared (IRS) spectroscopy.  
Here, we extend this to a much larger sample of GOODS-N sources out to $z \la 3$.  
%We extend the comparison of IR luminosity determinations based on 24~$\micron$ photometry to that when additional 70~$\micron$ photometry is available.  
%presented in \citet{ejm09a} 
%This analysis is done for the entire GOODS-N field for sources having spectroscopic redshifts out to $z\la3$.  
The redshift bins used, along with the number of sources per bin, per selection criteria, are given in Table \ref{tbl-2}.  
Only sources having spectroscopic redshifts are considered.  
Through this comparison, discrepancies between 24~$\mu$m-derived and best-fit IR luminosities are investigated.
Although the 70$\mu$m data are less sensitive than the 24~$\mu$m data, they help identify biases in the bolometric corrections at the bright end of the luminosity function.

\subsection{Trends with Redshift}
%To see whether the same discrepancies exist between these bolometric corrections  as a function of redshift, as reported by \citet{ejm09a}, we plot the ratio of 24~$\micron$ derived to the best-fit IR luminosities against redshift in the right panel of Figure \ref{fig-6} for sources having $f_{\nu}(24~\micron) > 200~\mu$Jy.  
In Table \ref{tbl-4} we give the median ratio of the 24~$\mu$m-derived, 24~$\mu$m-corrected, and radio-derived IR luminosities (see $\S$A.3.1) to the best-fit IR luminosities per redshift bin along with measured dispersions.  % (see $\S$3 for descriptions of each IR luminosity determination).  
The 24~$\mu$m-derived IR luminosities appear to do well by matching our best-fit IR luminosities up to redshifts of $z\la1.3$, at which point the 24$\mu$m-derived IR luminosities appear to overestimate the true IR luminosity by $\ga30$\%.    
More interestingly, we find that the dispersion in the ratios jumps significantly once past a redshift of $z\ga1.3$, increasing by a factor of $\sim$2 and 3 in the last two redshift bins, respectively.  

If we only consider those sources detected at 70~$\mu$m (i.e., the brightest sources) we find that the discrepancy between the 24~$\mu$m-derived and best-fit IR luminosities is even larger, along with the measured dispersions, once beyond $z\ga1.3$. 
%However, the discrepancy is even larger as 
The 24~$\mu$m-derived IR luminosity is a factor of $\sim$ 2 and 6 times larger than the best-fit IR luminosities in the two highest redshift bins.  
These discrepancies are similar to what was reported by \citet{ejm09a}, whose sample was selected to have $f_{\nu}(24~\micron) > 200~\mu$Jy and mid-infrared spectroscopy.      
This result is also consistent with that of \citep{cp07} whom found 24~$\mu$m derived IR luminosities to be a factor of 2 to 10 times larger than those derived from stacking 70 and 160~$\mu$m data for a sample of 24~$\mu$m bright (i.e., $f_{\nu}(24~\micron) > 250~\mu$Jy) galaxies lying in a redshift range between $1.5 \la z \la 2.5$.    
%are a factor of 2 to 10 lower than those inferred from using only 24~$\mu$m flux densities.
Consequently, it appears that luminosity, and not redshift, may be the more important parameter associated with the improper SED fitting when using 24~$\mu$m photometry alone.  
This result agrees with what was shown in $\S$\ref{sec-lir24corr} where the residual dispersion between the ratio of  the 24~$\mu$m-derived and best-fit IR luminosities was smaller when fitting versus luminosity as opposed to redshift. 

In comparing the radio-derived to best-fit IR luminosities among the 20~cm detected sources, we find that the radio (via the FIR-radio correlation) does not yield reliable IR luminosity determinations as they are $\ga$2 times larger, on average, in each redshift bin.  
This result is likely due to the difference in the depths of the radio and {\it Spitzer} surveys with the radio survey being much shallower, and thus less sensitive to galaxies of the same luminosity compared with the 24~$\mu$m data (see $\S$A.3.2).  

\subsection{Trends with 24~$\mu$m-Derived IR Luminosities}
In Table \ref{tbl-5} the same IR luminosity ratios are given as in Table \ref{tbl-4}, but per 24~$\mu$m-derived IR luminosity bin instead of redshift.  
%We find that that the 24~$\mu$m-derived IR luminosities are well matched to our best-fit IR luminosities for values of $L_{\rm IR}^{24} \la 10^{12}~L_{\sun}$, at which point the 24~$  \mu$m-derived luminosities begin to over predict the true values.  
For values of $L_{\rm IR}^{24} \ga 10^{12.5}~L_{\sun}$, the 24~$\mu$m-derived luminosity is a factor of $\sim$4 times larger than the best-fit values, on average.  
Not too surprisingly, when only considering those sources detected at 70~$\mu$m, we find nearly the same trend.  
A comparison between the radio-derived and best-fit IR luminosities among the 20~cm detected sources shows that for all luminosity bins where $L_{\rm IR}^{24} \ga 10^{10}~L_{\sun}$, the radio-derived luminosities are $\ga$2 times larger than the best-fit estimates.  
Interestingly, we find no trend with increasing luminosity bin.  
%the most discrepant bin is for values of  $10^{10}~L_{\sun} \la L_{\rm IR}^{24} \la 10^{10.5}~L_{\sun}$ where the radio-derived IR luminosity is $\sim$8 times larger than the best-fit estimate.  

%We also note that there are a number of sources which have high ratios (i.e., > 3) near $z\sim1$.  
%These sources are not detected at 16 or 70~$\micron$.  
%Accordingly, the high ratios likely arise from using the 16~$\micron$ upper limit to reject SEDs which were fit to the 24~$\micron$ data alone.  

\subsection{Comparison with Radio-Derived SFRs: The FIR-Radio Correlation versus $z$}
%\subsection{Radio Continuum Imaging}
The GOODS-N field has been imaged at 1.4~GHz using the Very Large Array for a total of 165 hours in all 4 configurations \citep{glenn10}.  
Since the VLA is centrally condensed, the observing time per configuration was scaled as follows, A-array (1), B-array (1/4), C-array (1/16), and D-array (1/64) \citep{fo08}.  
Using such an integration scaling provides the best sensitivity for extended sources.  
The final radio mosaic has a local RMS of $\sim$3.9~$\mu$Jy near the phase center covering the GOODS-N ACS area.  

Taking the published GOODS-N 1.4~GHz catalog of \citet{glenn10}, where 1230 discrete sources have been detected above a 5$\sigma$ threshold, a total of 342 of lie within the ACS$+$24~$\mu$m coverage. 
Of these sources, 286  could be matched to 24~$\micron$ detections.  
Among these, 194 had spectroscopic redshifts, and 192 were at $z<3$.   
The 20~cm flux densities among these 192 sources span a factor of $\sim$33, ranging from $\sim$$21.3-704~\mu$Jy.  
The median 20~cm flux density is $\sim$46~$\mu$Jy.  
Sources not matched with a radio counterpart were assigned the upper limit value of 19.5~$\mu$Jy.  

\begin{deluxetable*}{ccccccccc}
\tablecaption{Bolometric Correction Adjustments per $z$ Bin \label{tbl-4}}
\tablewidth{0pt}
\tablehead{
\colhead{} &
\multicolumn{2}{c}{} & 
\multicolumn{2}{c}{} & 
%\multicolumn{2}{c}{$f_{\nu}(24~\micron) >$ 200~$\mu$Jy} & 
\multicolumn{2}{c}{70~$\micron$ detected} & 
\multicolumn{2}{c}{20~cm detected}\\
\cline{6-9}\\
%\multicolum{2}{c}{\tableline} &
%\multicolum{2}{c}{\tableline} &
%\multicolum{2}{c}{\tableline}\\
\colhead{} &
 \colhead{$L_{\rm IR}^{24}/L_{\rm IR}$} & \colhead{$\sigma$} &
\colhead{$L_{\rm IR}^{\rm 24,corr}/L_{\rm IR}$} & \colhead{$\sigma$} &
%\colhead{$L_{\rm IR}^{16,24,70}/L_{\rm IR}$} & \colhead{$\sigma$} &
\colhead{$L_{\rm IR}^{24}/L_{\rm IR}$} & \colhead{$\sigma$} &
\colhead{$L_{\rm IR}^{\rm RC}/L_{\rm IR}$}& \colhead{$\sigma$}
}
\startdata
 0.0$~\leq~z~<~$ 0.4&  1.17&  0.33&  1.17&  0.33&  1.22&  0.23&  1.87&   2.38\\
 0.4$~\leq~z~<~$ 0.7&  1.08&  0.43&  1.08&  0.43&  1.10&  0.15&  1.67&   1.75\\
 0.7$~\leq~z~<~$ 1.0&  1.03&  0.28&  1.03&  0.27&  1.36&  0.25&  1.72&   3.20\\
 1.0$~\leq~z~<~$ 1.3&  1.19&  0.48&  1.17&  0.44&  1.36&  0.59&  2.76&  27.80\\
 1.3$~\leq~z~<~$ 1.9&  1.28&  0.96&  1.00&  0.26&  2.17&  1.39&  1.87&   3.10\\
 1.9$~\leq~z~<~$ 2.8&  1.56&  1.64&  1.00&  0.35&  6.06&  1.34&  3.25& 35.87
\enddata
\tablecomments{Only sources having spectroscopic redshifts considered.}
\end{deluxetable*}

\begin{deluxetable*}{ccccccccc}
\tablecaption{Bolometric Correction Adjustments per
  24~$\micron$-Derived IR Luminosity Bin \label{tbl-5}}
\tablewidth{0pt}
\tablehead{
\colhead{} &
\multicolumn{2}{c}{} & 
\multicolumn{2}{c}{} & 
%\multicolumn{2}{c}{$f_{\nu}(24~\micron) >$ 200~$\mu$Jy} & 
\multicolumn{2}{c}{70~$\micron$ detected} & 
\multicolumn{2}{c}{20~cm detected}\\
\cline{6-9}\\
%\multicolum{2}{c}{\tableline} &
%\multicolum{2}{c}{\tableline} &
%\multicolum{2}{c}{\tableline}\\
\colhead{} &
 \colhead{$L_{\rm IR}^{24}/L_{\rm IR}$} & \colhead{$\sigma$} &
\colhead{$L_{\rm IR}^{\rm 24,corr}/L_{\rm IR}$} & \colhead{$\sigma$} &
%\colhead{$L_{\rm IR}^{16,24,70}/L_{\rm IR}$} & \colhead{$\sigma$} &
\colhead{$L_{\rm IR}^{24}/L_{\rm IR}$} & \colhead{$\sigma$} &
\colhead{$L_{\rm IR}^{\rm RC}/L_{\rm IR}$}& \colhead{$\sigma$}
}
\startdata
$L_{\rm IR}^{24}~<~10^{10.0}$ &  1.19&  0.36&  1.19&  0.36&  1.22&  0.29&  3.01&   3.00\\
$10^{10.0}~\leq~L_{\rm IR}^{24}~<~10^{10.5}$ &  1.03&  0.38&  1.03&  0.38&  1.25&  0.12&  1.96&   3.75\\
$10^{10.5}~\leq~L_{\rm IR}^{24}~<~10^{11.0}$ &  1.00&  0.32&  1.00&  0.32&  1.12&  0.20&  2.29&  27.71\\
$10^{11.0}~\leq~L_{\rm IR}^{24}~<~10^{11.5}$ &  1.13&  0.42&  1.13&  0.42&  1.10&  0.12&  1.88&  23.22\\
$10^{12.0}~\leq~L_{\rm IR}^{24}~<~10^{12.5}$ &  1.69&  0.56&  1.00&  0.33&  1.50&  0.31&  1.83&   5.42\\
$L_{\rm IR}^{24}~\geq~10^{12.5}$ &  3.78&  1.96&  1.00&  0.38&  3.80&  1.58&  2.01&   3.71
\enddata
\tablecomments{Only sources having spectroscopic redshifts considered.}
\end{deluxetable*}

\begin{deluxetable*}{ccccccccc}
\tablecaption{IR-Radio Correlation per $z$ Bin \label{tbl-6}}
\tablewidth{0pt}
\tablehead{
\colhead{} &
\multicolumn{2}{c}{} & 
\multicolumn{2}{c}{70~$\micron$ detected} & 
\multicolumn{2}{c}{20~cm detected} &
\multicolumn{2}{c}{20~cm $+ 70~\micron$}\\
\cline{4-9}\\
%\multicolum{2}{c}{\tableline} &
%\multicolum{2}{c}{\tableline} &
%\multicolum{2}{c}{\tableline}\\
\colhead{} &
\colhead{$\tilde{q}_{\rm IR}$} & \colhead{$\sigma_{q}$} &
\colhead{$\tilde{q}_{\rm IR}$} & \colhead{$\sigma_{q}$} &
\colhead{$\tilde{q}_{\rm IR}$} & \colhead{$\sigma_{q}$} &
\colhead{$\tilde{q}_{\rm IR}$} & \colhead{$\sigma_{q}$}
}
\startdata
 0.0$~\leq~z~<~$ 0.4&  2.26&  0.39&  2.62&  0.28&  2.41&  0.27&  2.44&  0.19\\
 0.4$~\leq~z~<~$ 0.7&  2.17&  0.31&  2.49&  0.22&  2.42&  0.27&  2.46&  0.22\\
 0.7$~\leq~z~<~$ 1.0&  2.07&  0.36&  2.47&  0.16&  2.41&  0.26&  2.43&  0.12\\
 1.0$~\leq~z~<~$ 1.3&  2.10&  0.38&  2.50&  0.29&  2.22&  0.47&  2.38&  0.28\\
 1.3$~\leq~z~<~$ 1.9&  2.07&  0.29&  2.43&  0.16&  2.37&  0.31&  2.43&  0.18\\
 1.9$~\leq~z~<~$ 2.8&  2.10&  0.44&  2.66&  0.07&  2.13&  0.65&  2.66& 0.07
\enddata
\tablecomments{Only sources having spectroscopic redshifts considered.  
The logarithmic IR/radio ratios and associated dispersions are given in units of dex.}
\end{deluxetable*}

\subsubsection{Radio-Derived SFRs}
A nearly ubiquitous correlation is known to exist between the far-infrared (FIR; $42.5-122.5~\micron$) dust emission and predominantly non-thermal (e.g., 1.4~GHz) radio continuum emission arising from  star-forming galaxies \citep{de85, gxh85}.  
The most prominent feature of this correlation is that, at least in the local Universe, it extends for galaxies spanning nearly 5 orders of magnitude in luminosity while exhibiting a scatter which is less than a factor of $\sim$2 \citep[e.g.,][]{yun01}.  
Massive star formation provides the common link 
%between these two emission processes and appears to be the fundamental link 
relating these two emission processes both globally and on $\la$kpc scales within galaxy disks \citep[e.g.,][]{ejm06,ah06,ejm08}.  
Consequently, optically thin radio continuum emission is often considered to be a very good SFR diagnostic.  
While the FIR-radio correlation has largely been established in the local Universe, there are hints that it may remain constant out to high redshifts \citep[e.g.,][]{mg02,pa04,df06,ak06,ab06,cv07,ejm09a,ejm09b,ms10a,ms10b, nb10}.
%The difficulty with interpreting the results for a number of these studies arises from the lack of long-wavelength data and/or spectroscopic redshifts to properly constrain the full IR SED of each source allowing for a reliable comparison between the IR and radio continuum emission as a function of redshift, or too few galaxies to make statistical arguments for evolution or lack there of.  
%Now, having deep 70~$\micron$ imaging over the entire GOODS-N field, we can help improve this situation out to at least moderate redshifts.  

We parameterize the IR-radio correlation following a similar quantitative treatment of the FIR-radio correlation \citep[i.e.][]{gxh85}, except that we use the total IR ($8-1000~\micron$) luminosity, rather than the FIR fraction such that,    
\begin{equation}
\label{eq-qIR}
q_{\rm IR} \equiv \log~\left(\frac{L_{\rm IR}}{3.75\times10^{12}L_{\nu}(20~{\rm cm})}\right).
\end{equation}
Rest-frame 1.4~GHz radio luminosities are calculated such that, 
\begin{equation}
\label{eq-lrc}
L_{\nu}{(20~{\rm cm})} = 4\pi D_{\rm L}^{2} S_{\nu}(20~{\rm cm})(1+z)^{\alpha-1},
\end{equation} 
which includes a bandwidth compression term of $(1+z)^{-1}$ and a K-correction of $(1+z)^{\alpha}$ to rest frame 1.4~GHz.   
This assumes a synchrotron power law of the form $S_{\nu} \propto \nu^{-\alpha}$ with a spectral index $\alpha$ for which a value of  $\sim$0.8 is assumed \citep{jc92}.  
The median $q_{\rm IR}$ value reported for a sample of 164 galaxies without signs of AGN activity is 2.64 with a dispersion of 0.26~dex \citep{efb03}.  

By rewriting Equation \ref{eq-qIR}, we can use the radio measurements to independently obtain estimates of the IR luminosities of each source for comparison with our template fitting.  
%These 
For $q=2.64$ the radio-based IR luminosities are defined as 
%\begin{equation}
%\label{eq-l_IRrc_gen}
%L_{\rm IR}^{\rm RC} = 3.75 \times 10^{12+<q_{\rm IR}>}L_{\nu}(20~{\rm cm})
%\end{equation}
%or, by setting  $<q_{\rm IR}> \approx 2.64$, as
\begin{equation}
\label{eq-l_IRrc}
L_{\rm IR}^{\rm RC} = 1.64\times10^{15}L_{\nu}(20~{\rm cm})
\end{equation}
We assign a factor of $\sim$2 uncertainty to these luminosities since this is the intrinsic scatter among \citep[e.g.,][]{yun01,efb03} and within \citep[e.g.,][]{ejm06,ejm08} star-forming systems in the local Universe.  
%Radio derived SFRs, denoted as ${\rm SFR_{IR}^{RC}}$, are calculated using the values of $L_{\rm IR}^{\rm RC}$ along with Equation \ref{eq-SFRIR}.  

\subsubsection{Lack of Evolution with Redshift out to $z\la3$}
In Table \ref{tbl-6} we give the median IR/radio ratio for each subset of sources per redshift bin given in the top of Table \ref{tbl-2} along with the associated dispersion.   
By simply taking all 1107 24~$\micron$ detections with spectroscopic redshifts less than 3 we find median IR/radio ratios which are typically more than a factor of 3 lower than the locally measured value with a dispersion $\ga$ 2 for each redshift bin.  
This is likely driven by the fact that only 193 (i.e., $\sim$17\%) of these sources are radio detected while the remainder have IR/radio ratios calculated using the radio upper limits and are intrinsically much fainter.   
%their IR/radio ratios using upper limits for the radio flux densities and are intrinsically much fainter in the radio.  
%By instead computing IR/radio ratios for those sources having $f_{\nu}(24~\micron) > 200~\mu$Jy, such that the additional 70~$\micron$ data is helping to constrain the SED fitting and now only $\sim$50\% of the sources are not detected at 20~cm, we find the median IR/radio ratios to be within 1-$\sigma$ of the locally derived value for every redshift bin.  
%The dispersion also remains smaller than a factor of $\sim$2 for each redshift bin except for the highest bin, increasing to a factor of $\sim$2.34.  

Focussing on the subset of sources for which we have firm detections at 70~$\micron$, and are most confident in the determination of their estimated IR luminosities, we find that their IR/radio ratios also show no clear signs of evolution with redshift while displaying a dispersion that is a factor of $\la$2 for each redshift bin.  
In looking at the lowest redshift bin for these sources, we find the median IR/radio ratio and dispersion to be 2.62 and 0.28~dex, nearly identical to the average value of 2.64~dex  reported by \citet{efb03}.  
At the highest redshift bin, the median $q_{\rm IR}$ value also matches that for what measured in the local Universe.  
While the dispersion in $q_{\rm IR}$ among all 70~$\mu$m detected sources is found to be as small as what is measured in the local Universe, being $\sim$0.24~dex, we do note that the median value ($q_{\rm IR} = 2.50$~dex) is a factor of $\sim$1.37 (0.14~dex) lower than the value reported by \citet{efb03}. 
%somewhat lower, being (0.14~dex) lower than the value reported by \citet{efb03}. .    
%For this sample of galaxies we calculate a median value of $q_{\rm IR} = 2.50$~dex which is a factor of $\sim$1.37 (0.14~dex) lower than the value reported by \citet{efb03}.  
Since this departure is much smaller than the 1$\sigma$ dispersion in the locally measured IR-radio correlation, and smaller in magnitude compared to the errors associated with the SED template fitting, speculating on possible physical mechanisms driving this offset is not warranted.  
%While such a turn-up with increasing redshift may be expected, it is not statistically significant.  

Instead, taking the IR/radio ratios for only sources having 20~cm detections, we find that the median ratios are a factor of $\sim1.7-3$ times lower than what is measured in the local Universe.  
The dispersion in $q_{\rm IR}$ for these sources is also generally $\ga$2 in each bin, and increasing with redshift.  
%to a factor of nearly $\sim$4 in the highest redshift bins.  
As with the radio derived SFRs for all 20~cm sources, we attribute this systematic deviation in the IR/radio ratios to arise from the fact that the 20~cm detections are the brightest objects at each redshift for which the radio emission may have a considerable contribution from an AGN leading to overestimates in SFRs.    
%This may not be too surprising since we preferentially detect the brightest radio sources which may tend to be powered by AGN activity rather than star formation for which a low IR/radio ratio and increased dispersion is expected.  
Recent work by \citet{ms10a} has shown that this apparent negative evolution in the FIR-radio correlation arises from comparing IR and radio selected samples having dramatically different sensitivity limits, with IR observations typically being much deeper than data from current radio surveys.  
Taking this into account, \citet{ms10a} have used a survival analysis to demonstrate that the FIR-radio correlation out to similar redshift is consistent with the canonical value.  
Thus, we do not see any clear signatures of evolution in the FIR-radio correlation with redshift among the 70~$\micron$ detected sources in GOODS-N, nor any significant changes in the scatter per redshift bin, although the number of sources having spectroscopic redshifts for $z\ga1.3$ is highly limited.  
This is not surprising since any expected evolution with redshift due to increased energy losses to cosmic-ray electrons from inverse Compton scattering off of the CMB, resulting in depressed synchrotron emission from galaxies, should not become important until beyond $z\ga5$ \citep[e.g.,][]{ejm09b,bl10} 

%While the dispersion in $q_{\rm IR}$ among all 70~$\mu$m detected sources is found to be as small as what is measured in the local Universe, being $\sim$0.24~dex, we do note that the median value is somewhat lower.    
%For this sample of galaxies we calculate a median value of $q_{\rm IR} = 2.50$~dex which is a factor of $\sim$1.37 (0.14~dex) lower than the value reported by \citet{efb03}.  
%Since this departure is much smaller than the 1$\sigma$ dispersion in the locally measured IR-radio correlation, and smaller in magnitude compared to the errors associated with the SED template fitting, speculating on possible physical mechanisms driving this offset is not warranted.  

\section{Applicability of IRS Sample Results to the Full 24~$\mu$\lowercase{\rm m} Sample}
\begin{figure}
\begin{center}
\scalebox{0.5}{
\plotone{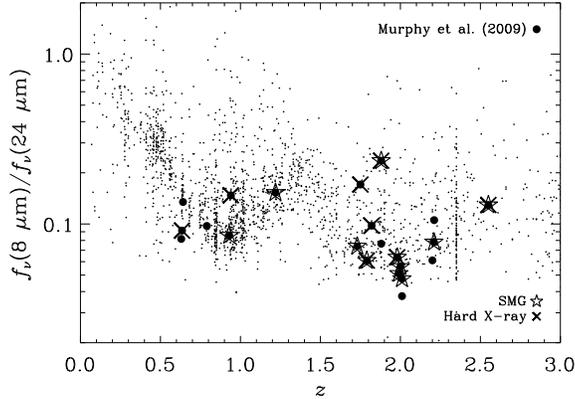}}
\end{center}
\caption{The ratio of 8 to 24~$\mu$m flux densities as a function of redshift for all 24~$\mu$m sources detected at 8~$\mu$m, and for which a redshift (spectroscopic or photometric) could be derived.  
Bold filled circles indicate the 22 galaxies included in the IRS study of \citet{ejm09a}; of these 22 galaxies, those which are SMGs and detected in the hard-band ($2.0-8.0$~keV) X-rays are identified by a star and cross, respectively.  
\label{fig-1A}}
\end{figure}

We compare the mid-infrared photometric properties of the 22 galaxies studied by \citet{ejm09a} to the entire 24~$\micron$-selected sample.  
This is done to demonstrate that applying the bolometric and AGN corrections derived in \citet{ejm09a} to the present sample of 24~$\micron$ sources is justified.  
In Figure \ref{fig-1A} we plot the observed ratio of 8 to 24~$\mu$m flux densities versus redshift for each 24~$\mu$m source detected at 8~$\mu$m, and for which a spectroscopic of photometric redshift could be measured (i.e., 2034 of 2196 24~$\mu$m detected sources in the ACS coverage).  
The 22 galaxies from \citet{ejm09a} are shown using bold filled circles.  
Sources which are SMGs, and presumably star formation dominated, are identified by stars while sources which are hard X-ray detected and likely harbor an AGN, are identified by crosses.  
We find that the sample of \citet{ejm09a} spans nearly all of the 8/24~$\mu$m phase space covered by the full 24~$\mu$m selected sample independent of the fact that the IRS sample was flux limited down to 200~$\mu$Jy.  

\begin{figure}
\begin{center}
\scalebox{0.6}{
\plotone{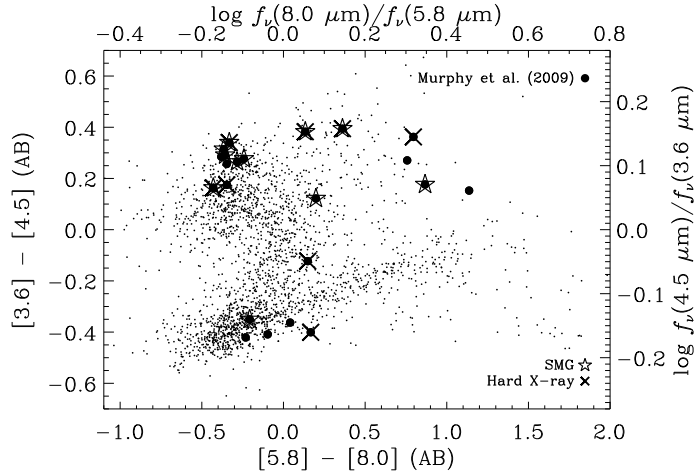}}
\end{center}
\caption{An IRAC color-color plot for all IRAC detected 24~$\mu$m sources usually used to identify galaxies hosting AGN \citep[e.g.,][]{ds05}.  
Bold filled circles indicate the 22 galaxies included in the IRS study of \citet{ejm09a}; of these 22 galaxies, those which are SMGs and detected in the hard-band ($2.0-8.0$~keV) X-rays are identified by a star and cross, respectively.  
The distribution of the IRS sample fills in most of the phase space covered in the IRAC color-color plot, indicating that the sample appears quite representative of galaxies spanning a range of types.  
\label{fig-2A}}
\end{figure}

A common diagnostic tool for identifying AGN is the IRAC color-color plot \citep[e.g.,][]{ml04,ds05}.  
While commonly used as a means to identify AGN, it is worth pointing out that this diagnostic does have its shortcomingings.  
Contamination of the AGN ``wedge" will occur by both low- and high-$z$ star-forming systems, which have similar IRAC colors \citep[e.g.,][]{pb06,jdon07,cc08}.  
In Figure \ref{fig-2A} we create such a plot for all IRAC detected 24~$\mu$m sources  (i.e., 2396 of 2664 24~$\mu$m detected sources) to see if the IRS sample only occupies a single region of this phase space.   
We again identify the sample of  \citet{ejm09a} using bold filled circles, while SMGs and hard-band ($2.0-8.0$~keV) detected sources are plotted using a star and cross, respectively.  
It is clearly shown that the IRS sample is well distributed in IRAC color-color space, touching locations occupied by AGN and star-formation dominated systems.  
Consequently, it appears that the IRS sample is representative of the entire 24~$\mu$m population, suggesting that applying the improved bolometric corrections and estimates for the fractional IR luminosity contributions from AGN  given in \citet{ejm09a} is warranted.   

\section{SFR Comparisons Among $B\MakeLowercase{z}K$ Sources: Occurrences of IR Excess Sources}
%\subsection{Occurrences of IR Excess Sources}
%\subsection{Among $BzK$ Sources: Occurrences of IR Excess Sources}
\citet{ed07a} have identified discrepancies between the UV, radio, and, IR derived SFR estimates in $z\sim2$ galaxies selected using the $BzK$ selection technique.
Following the criterion of \citet{ed07a,ed07b}, we calculate whether galaxies have a mid-infrared excess by measuring the ratio of IR$+$(observed) UV SFRs to that of the extinction corrected UV SFR.  
Sources having 
\begin{equation}  
\label{eq-irexc}
\log \left({\rm \frac{SFR_{IR} + SFR_{UV}}{SFR_{UV}^{corr}}}\right) > 0.5
\end{equation}
are considered to be ``mid-infrared excess" sources.  
We note that while the numerator contains a term for the UV emission which may escape the galaxy before being absorbed and re-radiated in the infrared by dust, this term is often negligible compared to the IR-based SFR term.  

\begin{figure}
\begin{center}
\scalebox{.50}{
\plotone{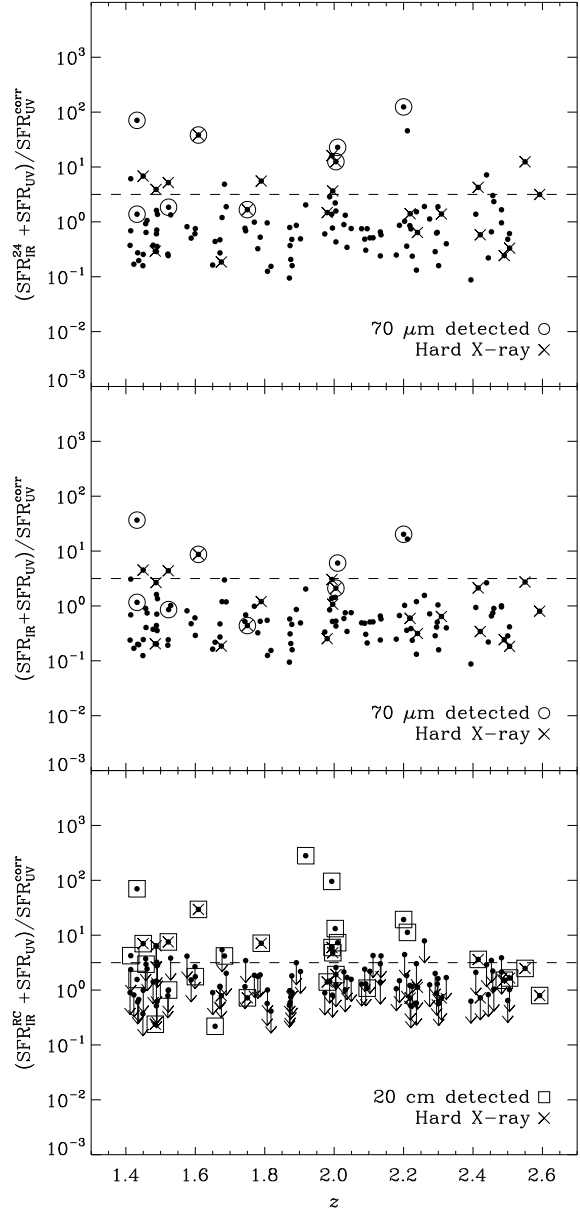}}
\end{center}
\caption{
We plot the ratio of IR-based SFR estimates  (plus the observed UV derived SFR) to extinction corrected UV SFRs as a function of redshift between $1.4 < z < 2.6$ for sources which we were able to measure rest-frame 1500~\AA~ flux densities.  
In the top panel the IR based SFR was calculated using IR luminosity estimates by SED fitting the 24~$\mu$m flux densities alone while the middle panel uses the best-fit IR luminosities to calculate the IR-based SFRs.  
In the bottom panel we use the 20~cm radio continuum flux densities and the FIR-radio correlation to derive the SFRs (${\rm SFR_{IR}^{RC}}$).
The SFRs are considered to be upper limits for those galaxies not detected at 20~cm (downward arrows).
%SFRs estimated by SED fitting 24~$\micron$ flux densities (plus a non-obscured UV derived SFR) to extinction corrected UV SFRs as a function of redshift.   
In each panel the dashed line indicates where the logarithm of this ratios is equal to 0.5~dex; galaxies having values higher than this are considered to be IR-excess sources \citep{ed07a}.   
%Galaxies having spectroscopic redshifts such that the rest-frame UV ($0.125 - 0.260~\micron$) slope could be fit by the $UBViz$ optical photometry are plotted; those without optical detections are shown with {\it downward arrows}.
Galaxies detected at 70~$\micron$ are identified with open circles while those that are detected in hard band ($2.0-8.0$~keV) X-rays are indicated by a cross.  
%The solid line indicates where the ratio between the ${\rm IR + UV}$ and UV corrected SFRs equal unity.  
%We plot the same quantities in the right panel, except that we use the 16, 24, and 70~$\micron$ photometry to derive the IR based SFRs (${\rm SFR_{IR}}$).
%The IR based SFRs are considered to be upper limits for those galaxies not detected at 70~$\micron$
%Same as Figure \ref{fig-6}, except that we use the 20~cm radio continuum flux densities and the FIR-radio correlation to derive the SFRs (${\rm SFR_{IR}^{RC}}$).
\label{fig-3A}}
\end{figure}

\begin{figure}
\begin{center}
\scalebox{.50}{
\plotone{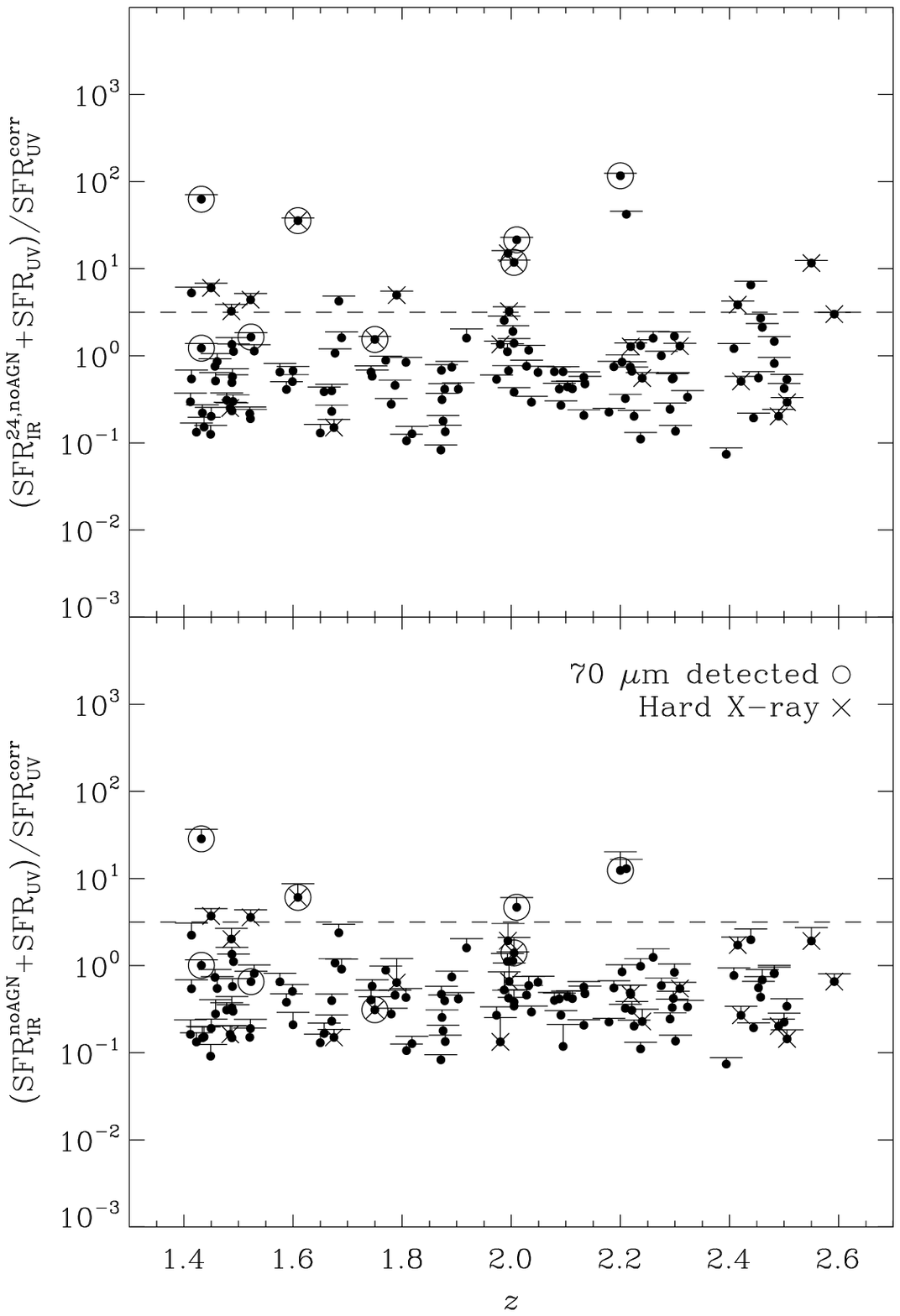}}
\end{center}
\caption{
The same as the top two panels in Figure \ref{fig-1A} except that we correct the 24~$\mu$m and best-fit IR luminosities for the contribution from AGN.  
The decrease is shown by a vertical line attached to a short horizontal line, and is 20\%, on average, in both panels.  
% for all galaxies having spectroscopic redshifts less than $z\$ and $f_{\nu}(24~\micron) > 200~\mu$Jy.  
The AGN luminosity used to correct the IR based SFR was determined using the relation derived in \citet{ejm09a} between the AGN contribution to the IR luminosity and the difference between best-fit IR luminosities and those derived from 24~$\micron$ photometry alone (see $\S$3.2.3).  %and those using our empirical correction (see $\S$\ref{sec-sed}).  %additional 16 and 70~$\micron$ data.
Sources that are detected in the hard band ($2.0-8.0$~keV) X-rays are identified by crosses.
% while those sources characterized as DOGs (Dey et al. 2008) are identified as {\it filled squares}.
%Those galaxies not detected in the optical, and whose UV SFRs were calculated using the upper limits of the optical photometry, are plotted with {\it downward arrows}.  
\label{fig-4A}}
\end{figure}

Using the 24~$\micron$-derived, radio-derived, and best-fit estimates for IR luminosity to derive SFRs, we plot the ratio of $\rm IR+UV$ to UV corrected SFRs for sources having spectroscopic redshifts in the range between $1.4 < z < 2.6$ detected at 24~$\micron$ for which we were also able to estimate a rest-frame 1500~\AA~ flux density.  
%optical photometry could be used to get an estimate of the rest-frame 1500~\AA~ flux densities of these sources.  
The IR based SFRs in the top panel of Figure \ref{fig-1A} are calculated using the 24~$\micron$-derived IR luminosities, the middle panel uses our best-fit IR luminosities, and the bottom panel uses IR-based SFRs estimated using the radio continuum imaging and the FIR-radio correlation.
%,  while those used to compute the IR-based SFRs in the bottom panel of Figure \ref{fig-1A} use the best-fit IR luminosities.  

Excluding the hard band ($2.0-8.0$~keV) X-ray detected sources, for which AGN are likely present, we find that the number of sources lying above the mid-infrared excess criterion decreases dramatically (i.e., from 7 to 4, a factor of $\sim$2) by using the best-fit IR luminosities rather than those derived using the 24~$\mu$m photometry alone.   
This is consistent with the findings of \citet{ejm09a} who reported that $\approx$50\% of their ``mid-infrared excess" sources could be accounted for by better constrained bolometric corrections rather than the subtraction of emission from obscured AGN which was estimated using deep mid-infrared spectroscopy.    
%This is not to say that these sources do not indeed host AGN
Instead, looking at the radio derived IR luminosities for those sources detected at 20~cm, we find that the number of mid-infrared excess sources is actually larger than when using the 24~$\mu$m derived IR luminosities, being 9 and 6, respectively.  
%similar to when using the 24~$\mu$m-derived IR luminosities, being 16 and 15, respectively.  

%We find that the number of sources lying above the mid-infrared excess criterion decreases dramatically (i.e., from 17 to 7, a factor of $\sim$2.5) by using the best-fit IR luminosities rather than those derived using the 24~$\mu$m photometry alone, consistent with the findings of \citet{ejm09a} who reported that $\approx$50\% of their ``mid-infrared" excess sources could be accounted for by better constrained bolometric corrections.  
%Instead, looking at the radio derived IR luminosities for those sources detected at 20~cm, we find that the number of IR excess sources is similar to when using the 24~$\mu$m-derived IR luminosities, being 16 and 15, respectively.  

\begin{figure*}
\plottwo{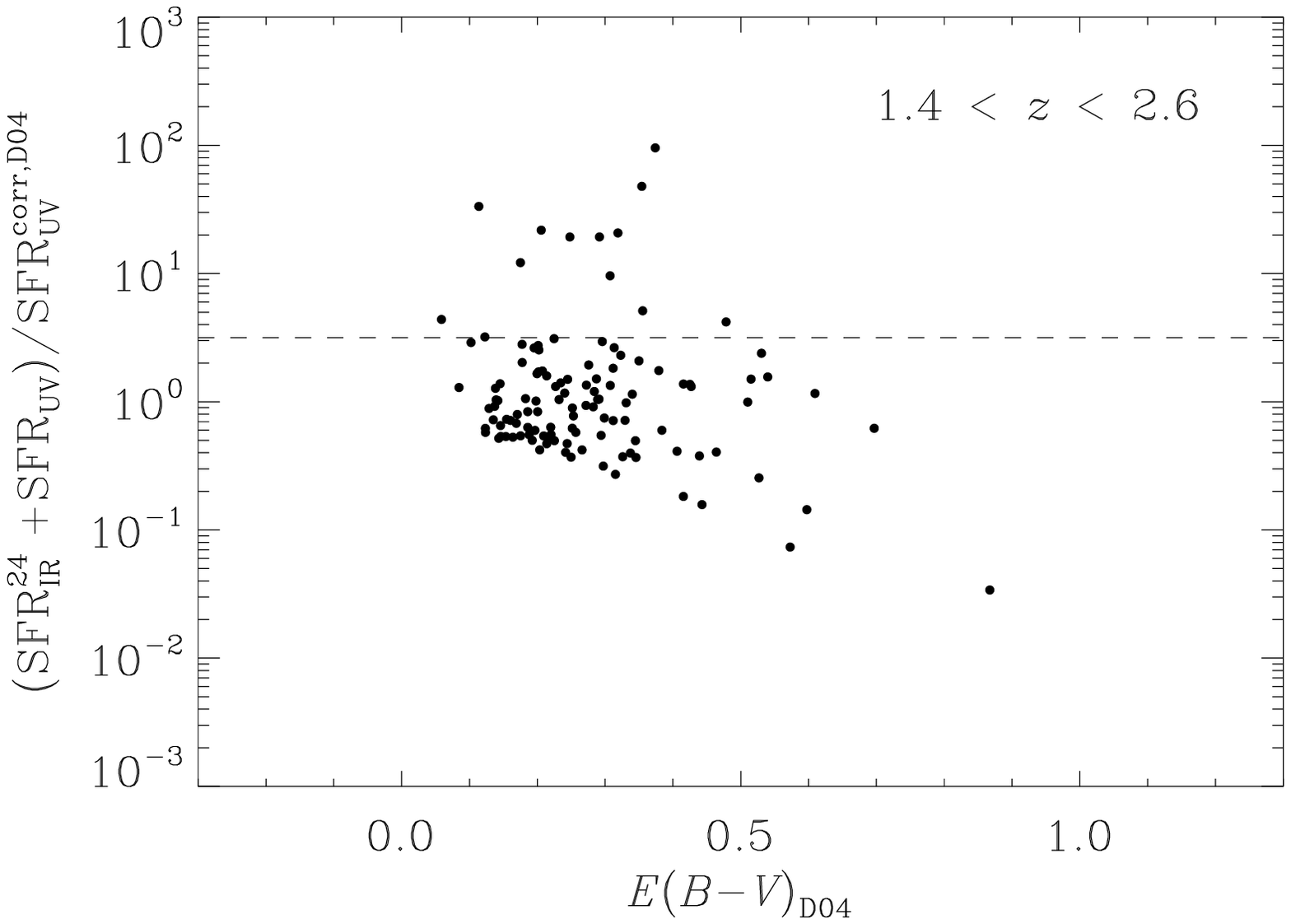}{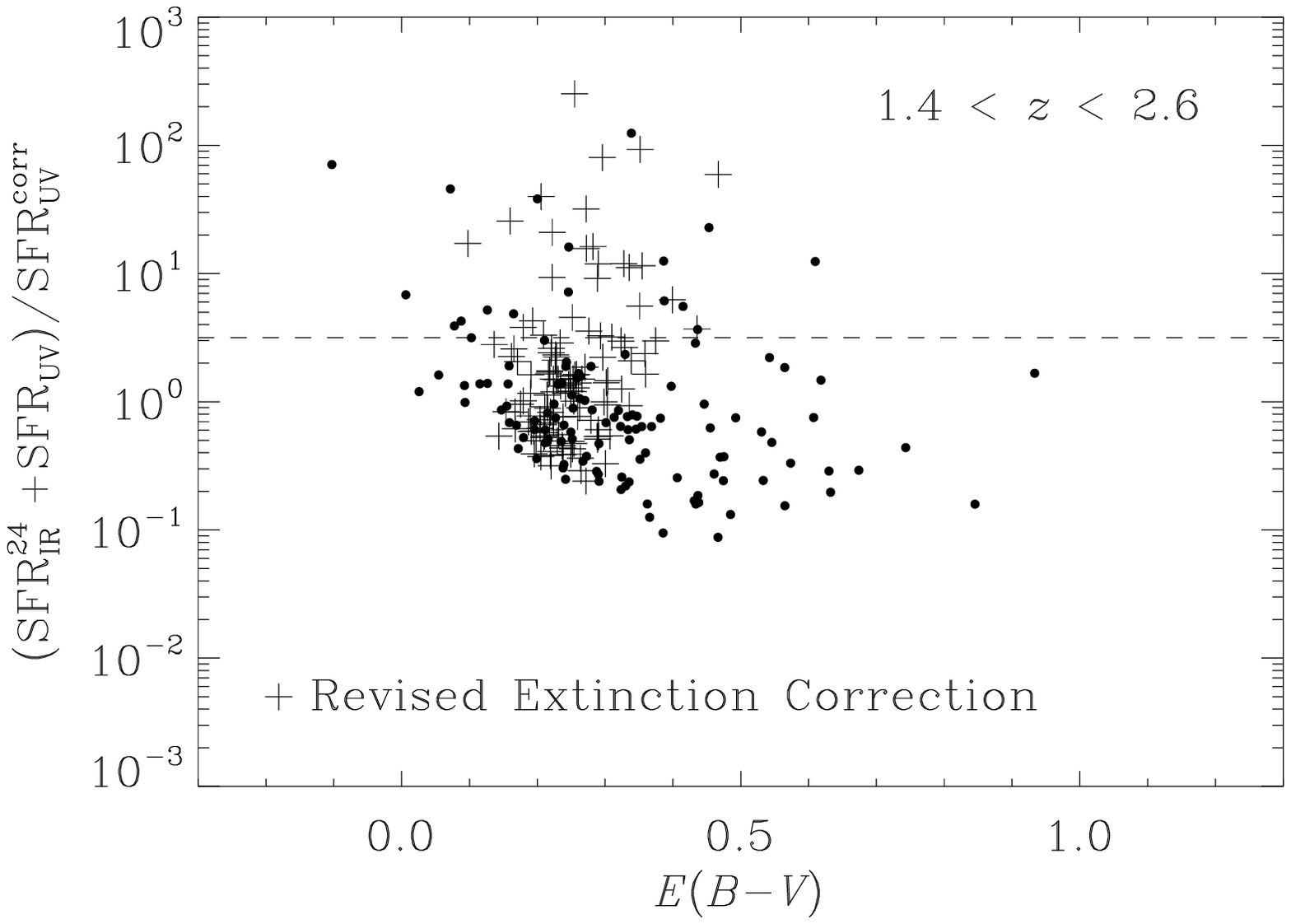}
%\plottwo{f11c_A2.eps}{f11d_A2.eps}
%\plottwo{f5e.eps}{f5f.eps}
\caption{ 
%We plot the
The ratio of IR derived SFRs estimated from SED fitting (plus UV derived SFRs uncorrected for extinction) to extinction corrected UV SFRs against the color excess $E(\bv)$ inferred from fitting the rest-frame UV ($1250 - 2600$~\AA) slope using $UBViz$ optical imaging data   
%In the top two panels we plot
for only those spectroscopically detected galaxies lying within a redshift range between $1.4 < z < 2.6$ (i.e., approximately the $BzK$ selection range).    
This is done to illustrate the difference in extinctions derived by using color relations (left panel; i.e., \citet{ed04}) to that when a proper slope is fit (right panel).  
The plus symbols in the right panel were generated using our revised extinction corrections (see $\S$4).  
%In the bottom panels this is done for all galaxies having spectroscopic redshifts for which fitting the rest-frame UV continuum was possible; the minimum redshift among these galaxies is $z \approx 0.66$.  
%In the bottom left panel, IR SFRs were derived from SED fitting 24~$\micron$ flux densities alone while, for the bottom right panel, IR based SFRs were derived from SED fitting  the 16, 24, and 70~$\micron$ photometry; those galaxies which are 70~$\micron$ detected are identified with circles.  
\label{fig-5A}}
\end{figure*}

\begin{figure}
\begin{center}
\scalebox{0.5}{
\plotone{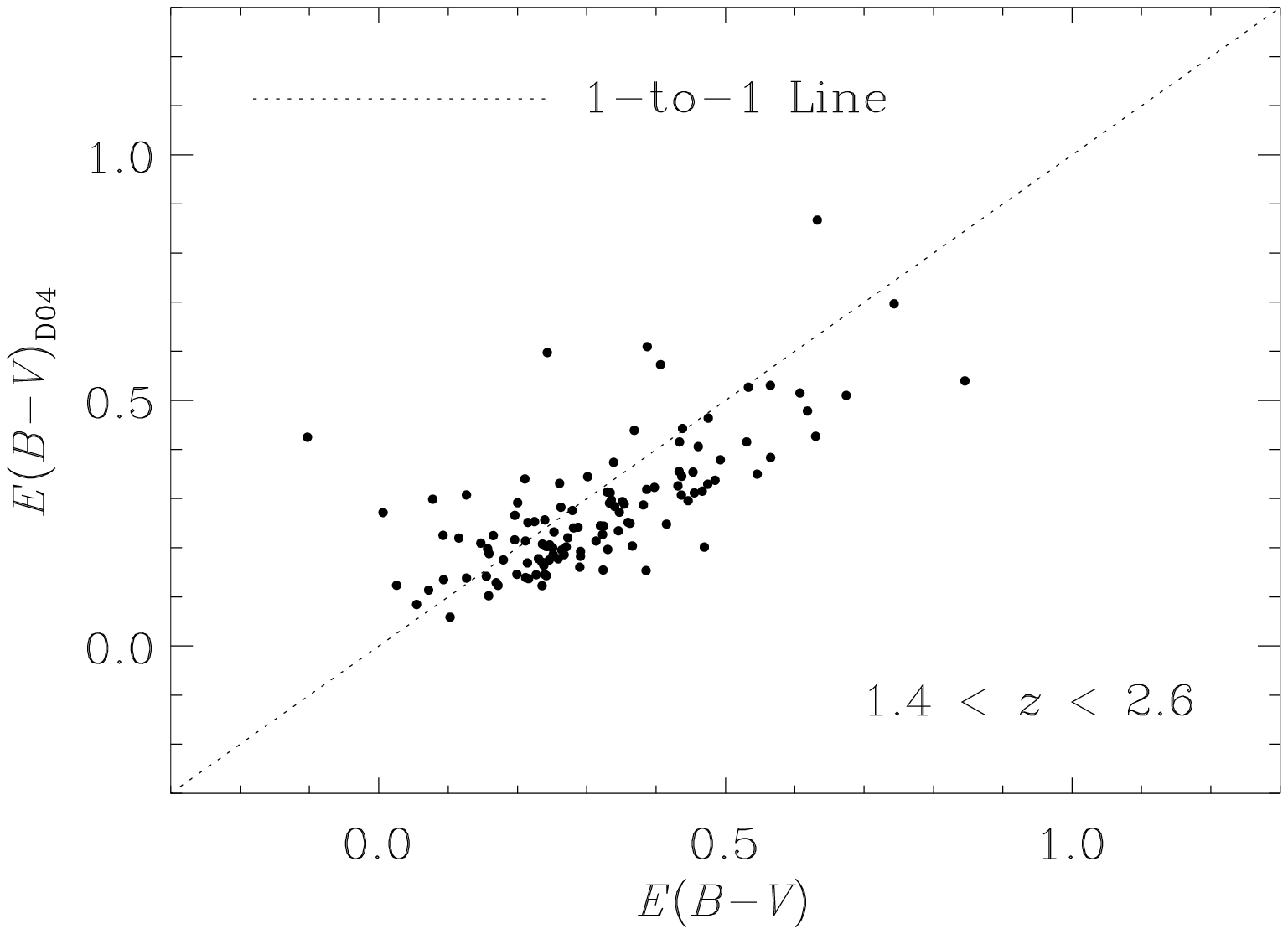}}
\end{center}
\caption{A comparison of extinction estimates for galaxies lying within a redshift range between $1.4 < z < 2.6$ (i.e., approximately the $BzK$ selection range).   
Extinctions calculated using color relations from \citet{ed04} are plotted against estimates from directly fitting the UV slope (see Figure \ref{fig-3A}).  
A one-to-one line is plotted as a dotted line.  
The direct fitting methods tends to yield larger reddening values.  
\label{fig-6A}}
\end{figure}

\subsection{Corrections for AGN}
%In $\S$ 4.1.3 
%We showed that the number of mid-infrared excess galaxies
%, being galaxies in a redshift range between $1.4 < z < 2.6$ whose ratio of IR (plus the observed UV) to extinction corrected UV SFRs is greater than a factor of $\approx$3, 
%is reduced by a factor of $\sim$2 by using our improved bolometric corrections rather than SED fitting 24~$\mu$m photometry alone.  
While the improved bolometric correction was able to account for $\approx$50\% mid-infrared excess sources (Figure \ref{fig-3A}), we find that by subtracting our estimates for the AGN contribution to the IR luminosities, all galaxies remain mid-infrared excess sources (top panel of Figure \ref{fig-4A}).  
The AGN contribution is indicated by the distance between the horizontal line and solid points.  
%only 2 of the mid-infrared excess sources are accounted for (top panel of Figure \ref{fig-2A}).    
This finding is consistent with \citet{ejm09a} who reported that the existence of the mid-infrared excess sources was dominated by an overestimate in the bolometric corrections rather than embedded AGN.   
This is contrary to the suggestion inferred by \citet{ed07a}, whose conclusion was based on the presence of a hard X-ray source through X-ray stacking.   
%Similar to \citet{ejm09a} we find that by using improved bolometric corrections and correcting for the presence of AGN, we are able to account for a substantial fraction of the sources which are identified to have a ``mid-infrared excess", as defined by \citet{ed07a},  based on their 24~$\micron$-derived IR luminosities (i.e., $\approx$50\%).   
%This indicates 
Accordingly, our result suggests that the sky and space densities of Compton-thick AGN reported by \citet{ed07b} are high by a factor of $\sim$2, and are likely more consistent with values of $\sim$1600~deg$^{-2}$ and $\sim$$1.3 \times10^{-4}$~Mpc$^{-3}$, respectively.  
% of mid-infrared excess sources.  }
%, as suggested by \citet{ed07a} based on stacking X-ray data of these galaxies which revealed the presence of a hard X-ray source.  }
% suggestive of obscured AGN activity.  

In the bottom panel of Figure \ref{fig-4A} we plot the ratio of AGN corrected IR (plus the observed UV) SFRs to extinction corrected UV SFRs and find the same occurrence  of mid-infrared excess sources (4 sources, 3 of which are 70$\mu$m detected), again suggesting that AGN are not causing the discrepancy between the IR and UV SFRs; 
among all the galaxies plotted, the application of Equation \ref{eq-agnfrac} suggests that AGN only contribute $\la$25\% of their total IR luminosity, on average.  
To explain the galaxies which persist as mid-infrared excess sources, \citet{ejm09a} suggested that it may have to due with improper extinction corrections relying on the slope of the UV continuum such that the 
%be more likely that these sources remain due to improper extinction corrections relying on the slope of the UV continuum; 
extinctions for these sources are underestimated, thereby yielding underestimates for the UV-derived SFRs.     
%This may not be too surprising given that there are catastrophic failures using the rest-frame UV slope to derive the extinction for galaxies having both low and high extinctions.  
In this scenario the extinction within a galaxy is so high that the ISM becomes optically thick, causing the relation between extinction and the slope of UV continuum to fail.  

%\subsubsection{Appropriateness of UV Extinction Corrections}
%{\bf KEEP DISCUSSION ABOUT EXTINCTION CORRECTIONS?}\\
\subsection{UV Extinction Corrections}
To compare how well simple color relations work to estimate extinction versus actually fitting the UV slopes of each objects, we plot the ratio of IR$+$UV to UV corrected SFRs in the in Figure \ref{fig-5A} against $E(\bv)$ color excesses calculated using the color relation given by \citet[][$E(\bv)_{\rm D04} = 0.25(B-z+0.1)_{AB};$ left panel]{ed04} and from a proper fit to the rest-frame UV continuum (right panel).  
%We also derive extinction corrections using a much simpler color relation given in \citet{ed04} where 
%\begin{equation}
%\label{eq-ebv_ed}
%E(\bv)_{\rm D04} = 0.25(B-z+0.1)_{AB}; 
%\end{equation}
%this equation was derived specifically for $BzK$ galaxies which span a redshift range of $1.4 < z < 2.5$.  
%Only galaxies in a redshift range between $1.4<z<2.6$ are considered as this is the range for which the color relation is considered to be valid.  
The UV corrected SFRs used in the ratios were calculated using the respective methods to compute the extinction corrections.  
Figure \ref{fig-6A} also compares the color excess $E(\bv)$ derived by the two methods.   
The direct fitting method tends to yield larger reddening values, particularly for objects with redder UV spectral slopes.  
The median $E(\bv)$ changes from $\sim$0.25 to 0.29 mag between the color-fitting $BzK$ method of \citet{ed04} and the direct fits to the UV slope for individual galaxies as defined here.   
The median UV extinction correspondingly changes from $\sim$2.0 to 2.4 mag, corresponding to an increase in the median UV-derived SFR by a factor of $\sim$1.5.    
Comparison of the two methods as a function of redshift shows  that the dispersion between them increases at the extremes of the redshift range for $BzK$ selection (i.e., near $z = 1.4$ and 2.6).    
For comparison, the sources in the right panel of Figure \ref{fig-5A} are re-plotted using the revised extinction corrections (i.e., Equation \ref{eq-irxb}) as plus symbols.    
%We find that the $E(B-V)$ color excess is $\sim$25\% larger, on average, %(i.e., UV corrected SFRs which are $\sim$50\% larger, on average) 
%when fitting the UV continuum as compared to using the color relation.  
%We also find that the dispersion in the color excess estimate is $\sim$25\% larger when the UV continuum is fit relative to when the color relation is used.  
Accordingly, using this color relation, rather than directly fitting the UV continuum for computing extinction corrections in this redshift range, does appear to introduce a systematic uncertainty for computing the UV corrected SFRs.   

%While no appreciable trends between the occurrences of IR excess sources and the derived extinction estimate are found when taking the subset of $1.4<z<2.6$ galaxies in the top two panels of Figure \ref{fig-3A}, this is clearly not the case when including the remainder of galaxies at $0.66 \la z \la 1.4$ (bottom panels of Figure \ref{fig-3A}).  
%For many of these galaxies, using their UV slopes to calculate the amount of internal extinction results in overestimates.  

%Only $\sim$2\% of sources for which the estimated color excess $E(\bv)>0.4$ are considered to be IR excess sources be the Equation \ref{eq-irexc} criterion, while $\sim$20\% of all sources with an $E(\bv)<0.4$ are identified as IR excess objects.  
%Consequently, whether a source is deemed to be an IR excess object appears to strongly correlate with the amount of the estimated extinction in the UV.    

%\clearpage

\end{document}